\documentclass{paper_format}
\usepackage{graphicx}
\usepackage{tabularx}
\usepackage{array}
\usepackage{multirow}
\usepackage{xcolor}
\usepackage[normalem]{ulem}
\useunder{\uline}{\ul}{}
\usepackage{hyperref}

\begin{document}
\selectlanguage{english}
\frenchspacing


\title{A Joint Synthetic Housing-Household Inventory}





\author{
\bf{Xiao Qian}%
\footnote{Department of Civil, Construction and Environmental Engineering, University of Delaware, Newark, DE 19716, USA}, 
\bf{Shangjia Dong}%
\footnote{Corresponding author, Assistant Professor, Department of Civil, Construction and Environmental Engineering, University of Delaware, Newark, DE 19716. sjdong@udel.edu},
\bf{Rachel Davidson}%
\footnote{Professor, Department of Civil, Construction and Environmental Engineering, University of Delaware, Newark, DE 19716}
}

\maketitle
\noindent

\begin{abstract}
\foreignlanguage{english}{%
\textbf{Abstract}

Accurately understanding the interactions between humans and the built environment requires integrated representations of both the buildings and the populations that occupy them. However, high-fidelity datasets that jointly capture detailed housing structures and demographic characteristics at the household level do not currently exist. This paper presents a framework for constructing a joint housing-household inventory that explicitly links individuals and households to compatible housing units from the National Structure Inventory (NSI), while preserving realistic population densities and demographic distributions. The framework integrates three components: (i) synthetic population generation from American Community Survey (ACS) Public Use Microdata Sample (PUMS) records that preserve complex intra-household relationships; (ii) a deep contrastive learning model that quantifies housing-household compatibility; and (iii) a hierarchical optimization-based allocation procedure that enforces building-level capacity and block-group-level demographic constraints. The generated synthetic population attains high statistical realism relative to the census microdata, and the contrastive learning model identifies compatible housing-household pairs with high predictive accuracy. Applied to coastal North Carolina, evaluations at building, neighborhood, and regional scales show that the joint inventory matches block-group-level demographic distributions, reproduces observed spatial population patterns without systematic bias, and maintains consistent allocation quality across urban, suburban, and rural contexts. By enabling coupled household- and building-level analyses, the resulting inventory supports a broad range of applications, including disaster resilience planning, housing and affordability analysis, energy-use assessment, and public health research.

}
\end{abstract}

\section{Introduction}

People live within, and depend upon, the physical structures and infrastructure that shape their communities. Our understanding of the built environment has expanded dramatically over the past two decades through the proliferation of open data sources such as Google Street View, high-resolution satellite imagery, and publicly maintained geospatial databases. Today, national-scale inventories such as the National Structure Inventory \citep{nsi} and Microsoft Building Footprints \citep{microsoft_us_building_footprints} catalog tens of millions of individual buildings, while detailed transportation networks are available through the U.S. Department of Transportation and OpenStreetMap. In parallel, population-level sociodemographic statistics and sampled household microdata from the U.S. Census Bureau \citep{us_census_acs} have enabled the construction of synthetic populations \citep{farooq2013simulation, borysov2019generate, kotelnikov2023tabddpm, qian2025deep} at multiple geographic scales, ranging from census block groups and tracts to counties and metropolitan regions.

Despite these advances, a fundamental gap remains: while we know with increasing precision \emph{where} structures are located and \emph{how} they are connected, we know comparatively little about \emph{who} lives within each structure and \emph{how} these spaces are actually used. This disconnect, between detailed knowledge of the built environment and limited knowledge of the populations that inhabit it, partly reflects necessary privacy protections embedded in U.S. data systems. However, it also constrains our ability to capture household-level experiences and responses to policy interventions, environmental change, technological disruption, and disaster events.

The consequences of this gap surface whenever analysts must infer household-level needs from average building or neighborhood conditions. Disaster loss models may misestimate displacement and recovery requirements; evacuation planning may overlook clusters of households requiring assistance; housing affordability programs may misallocate subsidies because resident income and tenure are unknown at the building level; energy efficiency interventions may be targeted at structures without regard to whether the occupant households are positioned to benefit; and public health analyses may miss risks that arise from the joint effect of building characteristics and occupant attributes. Each of these failures traces back to the same root cause: an inability to resolve \emph{who lives in which structure and unit.}

Addressing this gap requires a joint housing-household inventory that explicitly links households to their corresponding dwellings in space, capturing \emph{who lives where} at the resolution of individual buildings. Such an inventory would couple the spatial precision of structural databases with the demographic richness of household microdata, enabling building-level analyses that no single existing data source can support. Ideally, a joint synthetic housing-household inventory should achieve both realism and accuracy to meaningfully represent how people interact with the built environment. (1) \textbf{Population realism.} While individual-level census data are not publicly available, aggregated sociodemographic statistics and microdata (i.e., sampled household records) provide the foundation for constructing a synthetic population that approximates the true one. The resulting individuals and households should preserve realistic relationships among variables (e.g., education-income correlation), reproduce population-level distributions (e.g., age, tenure, gender, income, family structure), and capture spatial heterogeneity (e.g., neighborhood-level variations in age or income). (2) \textbf{Realistic housing and household relationships.} The inventory must accurately reflect how household attributes align with housing characteristics. For instance, household size, income, or tenure type collectively influence occupancy of units with suitable numbers of bedrooms, lot sizes, or building types, beyond simply matching bedroom counts. Since no dataset directly links households to specific housing units, the model must learn and represent the latent dependencies between household characteristics and housing features to ensure physical and behavioral plausibility. (3) \textbf{Realistic housing-household integration.} The process of joining the synthetic population with the housing inventory should respect physical housing capacities, maintain consistency with demographic marginals, and maximize compatibility between households and their assigned units according to the learned relationships. Rather than a simple assignment, this integration should jointly optimize household synthesis and housing, household matching under shared spatial and behavioral constraints.

\uline{\textbf{Challenges}} Developing such a joint inventory is a complex and computationally demanding task. We have made initial progress by tackling two core challenges: achieving population realism and learning realistic housing-household relationships. We introduced a transfer learning-based deep generative framework built on Variational Autoencoders (VAEs) to synthesize realistic populations that remain accurate (in terms of marginal distributions) when aggregated to the census block group (CBG) level \cite{qian2025deep}. Additionally, we developed a deep contrastive learning approach to capture latent relationships between housing and household attributes using co-occurrence patterns from microdata \citep{qian2025dcl}. This approach models housing-household matching as a continuous degree of compatibility rather than a binary outcome (matched or not), thereby enabling more robust and adaptive integration between household and housing inventories. Nonetheless, several key challenges remain before such frameworks can be fully operationalized. 
\begin{itemize}\vspace{-4pt}
    \item \textbf{Variable mismatch.} Population datasets and housing inventories often contain different sets of attributes. A housing-household relationship model learned from American Community Survey (ACS) microdata cannot be directly applied to a housing inventory like NSI that omits certain variables. Furthermore, similar variables may be recorded differently across datasets: for instance, income is recorded as a continuous value in PUMS but as categorical intervals in ACS census aggregates; similarly, age appears as exact years in PUMS but as grouped bins (e.g., 5-year age groups) in ACS census aggregates, requiring harmonization before integration. \vspace{-6pt}
    \item \textbf{Spatial scalability.} Constructing a regional inventory involves millions of housing units and households across thousands of CBGs.  The computational burden scales rapidly with population size. There is a need for an efficient, scalable inventory joining pipeline.\vspace{-6pt}
    \item \textbf{Joint inventory validation.} Evaluating the realism and accuracy of the resulting inventory remains difficult. No established standards exist for assessing the joint quality of population synthesis, housing-household relationships, and their integrated assignments, as prior research has typically focused on only one of these dimensions.
\end{itemize}

\uline{\textbf{Contribution}} This paper extends our prior work on (i) synthetic population generation \citep{qian2025deep} and (ii) deep contrastive learning (DCL)-based housing-house-hold compatibility modeling \citep{qian2025dcl}, and develops an integrated pipeline for producing a spatially explicit joint housing-household inventory. The major contribution is the first joint synthetic housing-household inventory that explicitly models compatibility relationships between households and their housing units, relationships that existing inventories have either ignored or reduced to random or rule-based assignment, and embeds these learned relationships into a constrained allocation at building-level resolution. The reproducible pipeline comprises three modules. We adopt TabDiff, a state-of-the-art mixed-type diffusion model, as the generative engine for synthetic population synthesis because of its higher distributional fidelity over CTGAN and TVAE and its ability to jointly model continuous, categorical, and ordinal variables; applying it here required substantial preprocessing of ACS PUMS microdata and retraining on the study-area sample. We employ the DCL architecture from our prior work for compatibility scoring, retrained on inputs aligned with a new cross-inventory variable mapping that bridges ACS PUMS household attributes and NSI building characteristics, enabling the DCL model to transfer compatibility scores to an inventory it was never trained on. Finally, we formulate a hierarchical, multi-objective MILP optimization that simultaneously enforces building-level capacity, aligns block-group-level demographic distributions with ACS targets, and maximizes aggregate DCL-derived compatibility; to our knowledge, no prior joint inventory has incorporated a learned compatibility measure into the allocation objective. The cross-inventory mapping, the compatibility-aware allocation formulation, and the resulting joint inventory constitute the methodological advances of this paper, with TabDiff, DCL, and MILP serving as adopted tools adapted to support these advances. We validate the resulting inventory at the building, neighborhood, and regional scales.

\section{Literature Review}

Recognizing the importance of a joint synthetic housing-household inventory for enabling household-level understanding and plan- and policy-making, several studies have contributed to this growing area of research. 

Early work by \citet{harada2017projecting} used geospatial data to allocate households to housing units based on building specifications such as type and location. This approach improved the precision of initial household placement by restricting the search space to a smaller set of plausible candidate buildings (i.e., geographically and structurally compatible units); however, within this candidate set, the final allocation still relied on random assignment.

\citet{rosenheim2021integration} advanced this line of work by developing the tool, \textit{pyncoda} \cite{rosenheim2025pyncoda}, which can produce a more granular joint housing-household inventory, significantly improving disaster impact assessment. Their approach involved constructing four distinct datasets: housing units, address points (each representing a unique residential address), buildings, and critical infrastructure. A Monte Carlo Simulation (MCS) was used to assign address points to housing units, and unique Random Merger Orders (RMOs) were created to link the housing unit and address point inventories. Census Block IDs and RMOs were then used to connect household-level census data to specific address points, thereby integrating all inventories. SimCenter BRAILS++ further integrated \textit{pyncoda} to allow researchers to allocate probabilistic households in building inventories, narrowing the gap between physical damage and community resilience. Although \textit{pyncoda} improves data linkage and granularity, it continued to depend on random or rule-based assignment of households to housing units, as it lacks explicit consideration of the relationship between household attributes and housing unit features.

More recently, \citet{ye2024enhancing} proposed an advanced framework that integrates LiDAR (Light Detection and Ranging) data, Points of Interest (POI), and quadratic programming to enhance the spatial granularity of population data. Their model constructs a detailed population distribution by combining diverse data sources and optimization techniques, thereby offering a structured, data-driven alternative to purely random allocation. Nevertheless, this approach remains limited by its focus on physical constraints, such as building capacity and population density, without explicitly modeling the relationships between household characteristics and housing attributes. 

Collectively, these studies demonstrate the evolution of joint housing-household modeling from random to increasingly data-driven approaches. Yet, several key challenges persist: current methods lack (1) a synthetic population that authentically represents sociodemographic diversity and spatial realism, and (2) a mechanism for learning and applying the latent relationships between household characteristics and housing features to join the population and building inventories. As \citet{aerts2018integrating} emphasized, individual and family characteristics strongly influence housing decisions, and neglecting these interdependencies reduces model realism and validity. Capturing these housing-household relationships during the integration of population and building inventories is therefore essential for generating realistic and practically useful joint inventories.

Our research addresses these challenges by redefining how the inventory integration problem is approached. Rather than generating a synthetic population of equivalent size to the building inventory and merging them directly, we develop a hierarchical optimization framework that selects synthetic households from an oversampled pool. This framework ensures that multiple criteria are simultaneously satisfied, including building capacity, sociodemographic marginal distributions at the census block group (CBG) level, and optimal housing-household compatibility derived from statistical relationships learned through deep contrastive learning. By embedding population realism and integration accuracy into the optimization process, our approach ensures the resulting joint housing-household inventory is both valid and representative.

\section{Methodology}

The American Community Survey (ACS) Public Use Microdata Sample (PUMS) and National Structure Inventory (NSI) provide the empirical foundation for constructing the synthetic joint housing-household inventory. The ACS PUMS contains detailed, anonymized records for individual persons and housing units. For our experiments, we selected a representative set of household- and individual-level attributes to be included in the synthetic population, summarized in Table \ref{tab:inventory-attributes}. It is important to note that these choices are illustrative rather than restrictive. The proposed framework can synthesize any subset of attributes available in the ACS PUMS and NSI.

\begin{figure}[htbp]
    \centering
    \includegraphics[width=0.9\linewidth]{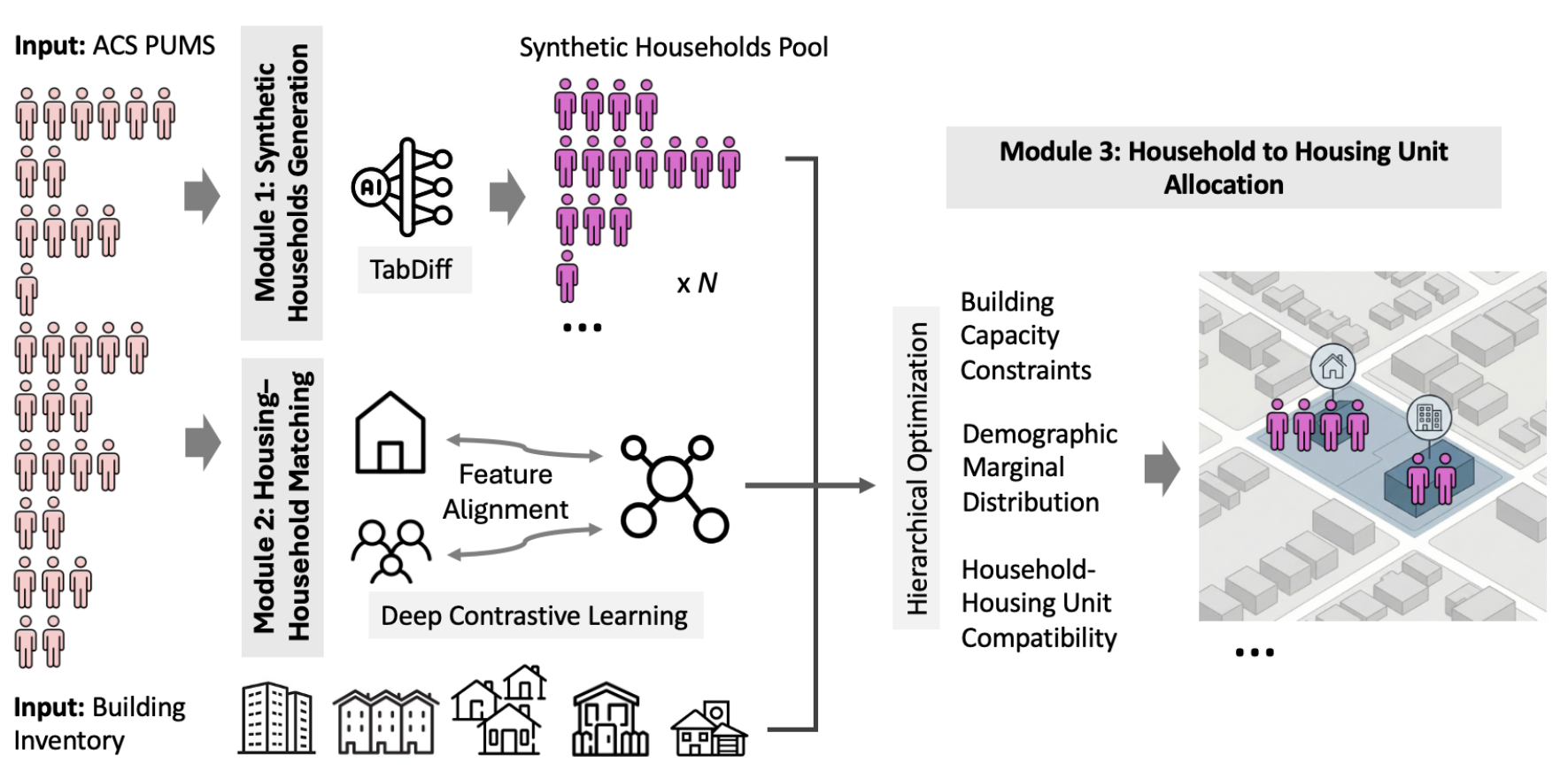}
    \caption{\textbf{Schematic overview of the proposed housing-household inventory joining framework.}}
    \label{fig:framework}
\end{figure}

Figure~\ref{fig:framework} presents an overview of the proposed integrative pipeline. Module 1 (Section~\ref{sec:synthetic_household_data_generation_model}) generates an oversampled, statistically realistic synthetic population of households and individuals using ACS PUMS data. Module 2 (Section~\ref{sec:housing-household_matching}) learns housing-household relationships and computes compatibility scores between the oversampled synthetic households produced in Module 1 and housing units within NSI buildings. Module 3 (Section~\ref{sec:Household-to-Housing-allocation}) allocates households to feasible housing units using a MILP-based optimization model, producing a spatially explicit joint inventory that links persons and households to housing units in different buildings.

\begin{table}[!ht]
    \centering
    \caption{Household, individual, and housing unit features included in the joint synthetic housing-household inventory.}
    \label{tab:inventory-attributes}
    \resizebox{\textwidth}{!}{%
    \begin{tabular}{>{\raggedright\arraybackslash}p{0.12\textwidth} 
                    >{\raggedright\arraybackslash}p{0.38\textwidth} | 
                    >{\raggedright\arraybackslash}p{0.12\textwidth} 
                    >{\raggedright\arraybackslash}p{0.38\textwidth}}
    \hline
    \multicolumn{2}{l|}{\textbf{Housing Unit Features}}                   & \multicolumn{2}{c}{\textbf{Synthetic Population}}                                                                                                                                                                                             \\ \hline
\textbf{Variable}    & \textbf{Description}                           & \textbf{Variable} & \textbf{Description}                                                                                                                                                                                                      \\
                     &                                                & \multicolumn{2}{c}{\textit{Household}}                                                                                                                                                                                                        \\ \hline
VALP                 & Property value. Numeric.                       & TEN               & \begin{tabular}[c]{@{}l@{}}Tenure. Categories are: 1:$\sim$Owned with \\ mortgage or loan, 2:$\sim$Owned free and \\ clear, 3:$\sim$Rented, and 4:$\sim$Occupied \\ without payment of rent.\end{tabular}                 \\ \hline
YRBLT                & Year structure built. Ordinal categories representing construction periods (e.g., 1939 or earlier, 1940--1949). & VEH               & \begin{tabular}[c]{@{}l@{}}Vehicles available. Categories range \\ from 0 (No vehicles) to 6 (6 or more \\ vehicles).\end{tabular}                                                                                        \\ \hline
ACR                  & Lot size. 1:$\sim$Less than one acre, 2:$\sim$1 to 10 acres, 3:$\sim$10 or more acres. & HHL               & \begin{tabular}[c]{@{}l@{}}Household language. 1:$\sim$English only, \\ 2:$\sim$Spanish, 3:$\sim$Other Indo-European \\ languages, 4:$\sim$Asian and Pacific Island \\ languages, or 5:$\sim$Other language.\end{tabular} \\ \hline
R65                  & Presence of persons 65 years and over. 0:$\sim$None, 1:$\sim$One person, 2:$\sim$Two or more. & HINCP             & \begin{tabular}[c]{@{}l@{}}Household income over the past 12 \\ months. Numeric.\end{tabular}                                                                                                                             \\ \hline
\multirow{3}{*}{BLD} & \multirow{3}{=}{Units in structure. 01:$\sim$Mobile home, 02--03:$\sim$Single-family, 04--09:$\sim$Multi-family, 10:$\sim$Other.} & \multicolumn{2}{c}{\textit{Individual}}                                                                                                                                                                                                       \\ \cline{3-4} 
                     &                                                & SEX               & Sex. 1:$\sim$Male or 2:$\sim$Female.                                                                                                                                                                                      \\ \cline{3-4} 
                     &                                                & AGEP              & Age in years. Numeric.                                                                                                                                                                                                    \\ \hline
DIS                  & Disability status. 1:$\sim$With a disability, 2:$\sim$Without a disability. & SCHL              & \begin{tabular}[c]{@{}l@{}}Educational attainment. 25 categories \\ range from ``No schooling completed'' \\ to ``Doctorate degree''.\end{tabular}                                                                            \\ \hline
    \end{tabular}
}
\end{table}

\subsection{Synthetic Households Generation Module }\label{sec:synthetic_household_data_generation_model}

We begin by filtering the ACS PUMS records to retain only occupied housing units (\texttt{TYPEHUGQ} = 1) with at least one resident (\texttt{NP} > 0). We then apply a targeted recoding to the educational attainment variable, \texttt{SCHL}, to align it with the spatial benchmarks used later in the pipeline. In ACS, \texttt{SCHL} is recorded as not applicable for children under age 3 and is defined as the highest degree or level of school completed for respondents age 3 and older. However, the small-area ACS education tabulations used for block-group benchmarking are reported only for the population age 25 and over \citep{acs_subject_definitions_2022, census_education_about_2026, census_b15003_2020}. For respondents younger than 25, \texttt{SCHL} therefore conflates ongoing school progression with terminal attainment and does not align with the adult-education marginals enforced elsewhere in the framework. We accordingly recode \texttt{SCHL} to a value of 0 for all individuals younger than 25, while retaining the original ACS codes 1-24 for respondents age 25 and over. Because the diffusion model treats \texttt{SCHL} as a categorical token rather than as an ordered numerical score, this preprocessing value functions as a distinct, age-dependent state representing education that is not yet comparable to adult attainment; it does not correspond to the substantive ACS category No schooling completed,'' which remains code 1. When the framework is transferred to a microdata source with a different definition of youth education, this recoding rule should be adjusted to follow the source-specific definition.

After data restructuring, we reorganized the household- and person-level records into a unified structure following the approach of \citet{qian2025deep}. Specifically, instead of storing each individual as a separate row, all person records within a household are flattened into a single row representing that household. This restructuring is essential: It enables the deep learning model to learn the joint distribution of household and individual characteristics, and to capture the higher-order dependencies among multiple individuals within the same household.

\paragraph{Generative Model Architecture and Details}

Our earlier work introduced an end-to-end population synthesis pipeline based on a transfer-learning-enabled VAE \citep{qian2025deep}. While it produces realistic households and the individuals within and preserves marginal sociodemographic distributions, two issues limit its suitability for statewide inventories: generation is too slow to scale to county- or state-level deployments, and tabular generators based on VAEs or GANs struggle to capture dependencies across heterogeneous fields, particularly when long-tail categorical features are paired with skewed continuous variables. Recent studies report that VAE-GMM hybrids occasionally collapse and that high-cardinality VAEs require heavy regularization to model cross-type dependencies \citep{apellaniz2024vaegmm, carlin2025cardicat}. These limitations are especially problematic for ACS microdata, where multi-person households must maintain coherence between household-level traits (e.g., tenure, language, mobility) and individual-level attributes (e.g., age, education).

\begin{figure}[ht]
    \centering
    \includegraphics[width=0.99\textwidth]{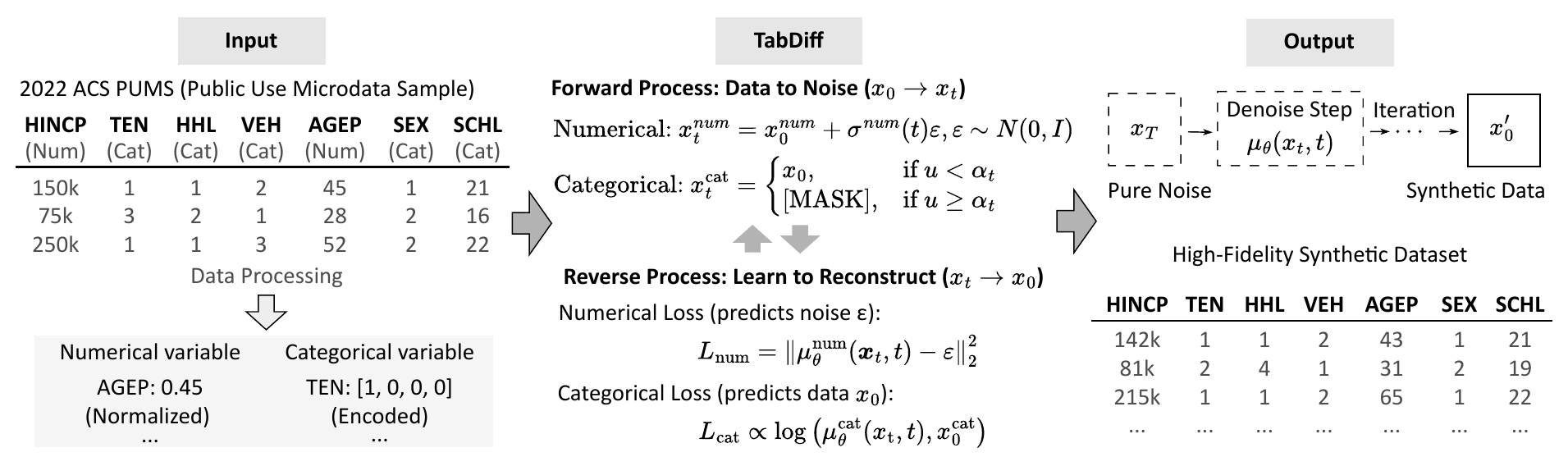}
    \caption{TabDiff architecture for synthetic household data generation. }
    \label{fig:tabdiff-architecture}
\end{figure}

Diffusion models have recently emerged as the state of the art for synthetic data generation, consistently outperforming leading GAN and VAE baselines on Fréchet Inception Distance (FID) and related fidelity metrics \citep{dhariwal2021diffusion, wang2024diffusionreview}. Several extensions adapt this approach to tabular data: STaSy showed that score-based diffusion can outperform CTGAN and TVAE on mixed-type data \citep{kim2023stasy}; TabDDPM improved discrete-variable handling via noise-schedule refinements \citep{kotelnikov2023tabddpm}; and TabSyn improved joint-distribution fidelity using latent diffusion with conditional alignment \citep{zhang2024tabsyn}. Building on these advances, TabDiff reports a 22.5\% improvement in pairwise correlation accuracy over prior tabular generators while enforcing semantic column constraints \citep{shi2025tabdiff}. We therefore adopt TabDiff as the generative engine. Its denoising network couples column-wise embeddings with a transformer to reconstruct continuous, categorical, and ordinal variables simultaneously within a single score model, making it well-suited to the heterogeneous structure of ACS household microdata. Figure~\ref{fig:tabdiff-architecture} illustrates the architecture and its three core components.

The \textbf{\textit{data preparation process}} prepares all the different types of household and individual information before they are used by the model. As shown in the upper-left part of Figure~\ref{fig:tabdiff-architecture}, each household is converted into a fixed-length sequence of tokens (i.e., a small unit of data that a model processes) that the model can understand. Numerical variables (e.g., income, age, vehicles) are normalized via z-scores, while categorical variables (e.g., sex, education) retain ACS coding and are embedded for model processing. We generate special masks to help the model understand that some tokens represent categories (e.g., sex, education) rather than numbers, so it should process them differently during training, especially during the noise-injection steps of the diffusion process. If a household includes multiple individuals, their information is added as extra tokens in the sequence. We also include padding masks, so the model knows which tokens represent actual individuals in a household and which tokens are just placeholders to keep all households at a fixed sequence length; this ensures the diffusion steps only modify the actual data and ignore the padded positions.

The \textbf{\textit{forward diffusion process}} adds noise to the original clean data in a way that respects the differences between numerical and categorical variables. For numerical attributes, noise is added gradually following $x_0$ according to $x_t^{\text{num}} = x_0^{\text{num}} + \sigma^{\text{num}}(t)\epsilon$, where is $\epsilon$ is the Gaussian noise $\mathcal{N}(0, I_{M_{\text{num}}})$. For categorical attributes, we use a masking-based strategy instead of adding continuous noise. At each timestep $t$, a category value is either kept as the original value or replaced with a special [MASK] token, depending on a probability determined by $\alpha_t$: 
\begin{equation}
    x_t^{\text{cat}} = \begin{cases} x_0, & \text{if } u < \alpha_t \\ [\text{MASK}], & \text{if } u \geq \alpha_t \end{cases}
\end{equation}
This ensures that categorical variables are ``corrupted'' in a way that fits their discrete nature. Together, these two approaches allow numerical and categorical variables to follow noise injection schedules that are appropriate for their data types, while still producing a consistent t=0 to fully corrupted data at timestep $T$.

The \textbf{\textit{reverse diffusion process}} uses a transformer-based denoiser $\mu_\theta$ equipped with residual attention blocks to recover clean data from noisy inputs. The denoiser is trained using two complementary objectives: a numerical reconstruction loss, 
\begin{equation}
    \mathcal{L}_{\text{num}} = \|\mu_\theta^{\text{num}}(x_t, t) - \epsilon\|_2^2
\end{equation}
and a categorical prediction loss, 
\begin{equation}
    \mathcal{L}_{\text{cat}} = \frac{\alpha_t'}{1-\alpha_t} \log\langle\mu_\theta^{\text{cat}}(x_t, t), x_0^{\text{cat}}\rangle
\end{equation}
The transformer shares parameters across all categorical groups but uses separate projection heads for each variable type (e.g., household-level categorical like tenure, individual-level categorical like sex, household-level numerical like income, individual-level numerical like age). Multi-head self-attention over these mixed-type tokens captures higher-order dependencies, enabling the generation of coherent multi-person households that preserve both statistical realism and logical consistency.

We adopt the official TabDiff PyTorch implementation, which provides configuration files for the learnable-schedule variant and reproducible training scripts for mixed-type tabular datasets \citep{tabdiff_github}. Our model is trained for 8,000 steps using a power-mean noise scheduler. The denoiser follows TabDiff's transformer-based design with two transformer layers, a hidden width of 1024, and a single attention head. We use the AdamW optimizer with a learning rate of $0.001$ and no weight decay. Validation is conducted every 2,000 steps, and we do not apply early stopping or gradient clipping. All experiments are executed in WSL 2 on Ubuntu 24.04.3 LTS, running on an Intel Core i7-13700KF (24 threads) and an NVIDIA GeForce RTX 4070 Ti (12 GiB; driver 581.15; CUDA 13.0). We use the official PyTorch CUDA builds with Python 3.12.3.

\subsection{Housing-Household Matching Module}\label{sec:housing-household_matching}

Random or capacity-only assignment of households to housing units \citep{rosenheim2021integration, ye2024enhancing} fails to capture who actually lives where. We therefore adopt the deep contrastive learning (DCL) housing-household compatibility model of \citet{qian2025dcl}, which provides the methodological foundation for learning compatibility scores from observed ACS housing-household pairs. Classical housing theory motivates such a learned approach: hedonic and residential sorting models describe households as choosing among differentiated bundles of dwelling and neighborhood attributes under budget constraints, heterogeneous preferences, and market conditions \citep{rosen1974hedonic, quigley1985consumer, bayer2004equilibrium}. Therefore, observed pairs reflect multidimensional sorting rather than isolated pairwise correlations. The DCL model recovers high-dimensional, nonlinear compatibility patterns from observed matches in ACS PUMS, including dependencies among income, tenure, household size, and structure type that are difficult to encode with hand-crafted rules. We accordingly represent suitability as a continuous compatibility score rather than a binary feasible/infeasible indicator: a binary formulation would treat, for instance, both three- and four-bedroom units as equally suitable for a four-person household, masking meaningful differences in fit and yielding suboptimal downstream allocations.

However, constructing such a measure is challenging. ACS PUMS \citep{us_census_acs}, the main publicly available source of housing-household relationships, has three critical limitations. First, it records only actual household-home pairs, so explicit negative examples are unavailable. An unobserved pair may simply be absent from the sample, which does not constitute evidence of incompatibility; this differs from many contrastive-learning settings, where non-co-occurring pairs can be treated more directly as negatives because semantic mismatch is clearer from the data structure. Second, although a given home could realistically fit many household types, only one observed pairing is recorded per unit, and inferring alternative matches through clustering can introduce errors. Third, household and housing data describe fundamentally different objects, people versus buildings, shaped by distinct processes (socioeconomic conditions versus housing development). Unlike standard multi-modal data such as paired image-text views of the same object \citep{huh2024platonic}, three persons in a household and three rooms in a unit are not equivalent even when they appear identical in tabular form; what links them is compatibility rather than identity, and mapping them into a single shared representation risks blurring this distinction. The DCL architecture introduced by \citet{qian2025dcl} addresses these challenges, and we adopt it as the compatibility component of the joint-inventory pipeline.

\paragraph{Deep Contrastive Learning Model}

\begin{figure}[t]
\centering
\includegraphics[width=\linewidth]{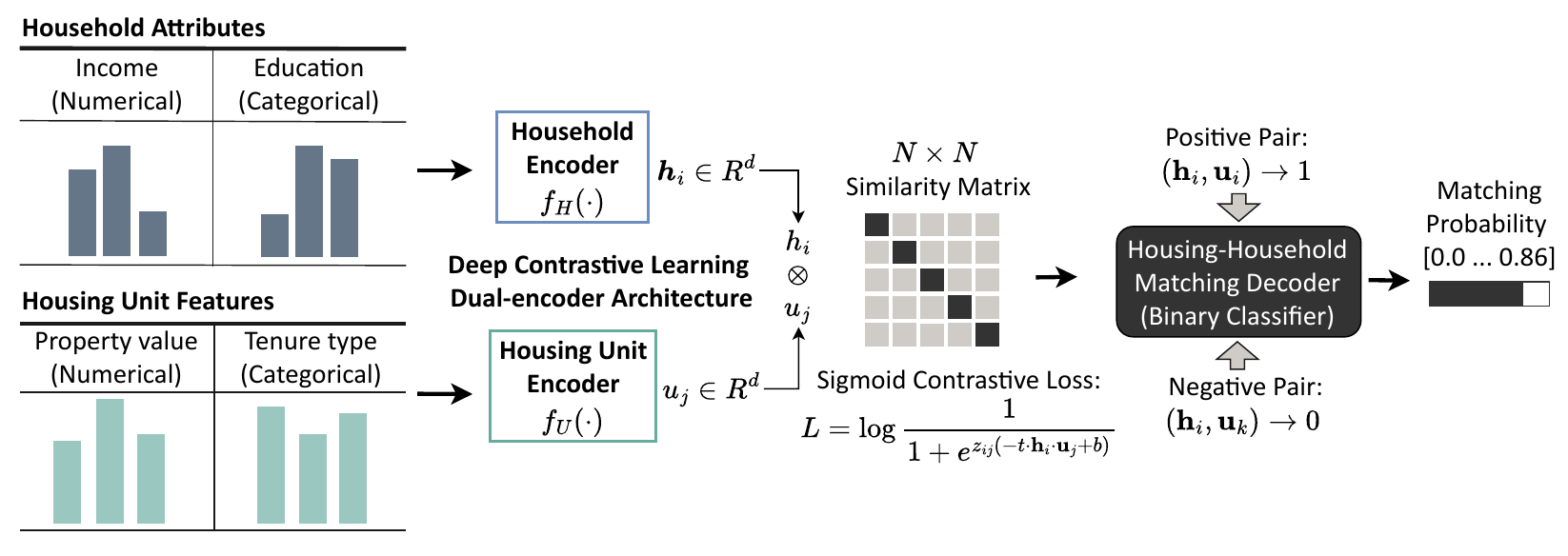}
\caption{\textbf{Contrastive learning framework for household-structure matching.} }
\label{fig:contrastive}
\end{figure}

The deep contrastive learning (DCL) model in Figure~\ref{fig:contrastive} learns housing-household compatibility by projecting household features (e.g., income, size, demographics) and housing-unit features (e.g., value, size, amenities) into a shared representation space, where compatible pairs are pulled closer together, and incompatible pairs pushed apart, producing a continuous matching score for each household-unit pair.

Because PUMS records only the structural attributes of the units households occupy, the model is trained in a self-supervised manner: observed pairs serve as positives, and unobserved pairs as negatives. To reduce noise from unobserved pairs that may nevertheless be plausible matches, households and housing units are first clustered by feature similarity, with 700 household clusters and 1,000 housing-unit clusters used in the North Carolina application. Pairs drawn from corresponding clusters are then treated as additional plausible positives. This strategy expands training coverage, improves robustness, and helps the model capture latent patterns in housing-household relationships.

The model uses a dual-encoder neural network with separate MLP encoders for households and housing units, preserving domain-specific features while aligning them in a shared latent space. Each encoder includes feature-specific embeddings for categorical and numerical attributes, a six-layer residual MLP with 2,048 hidden units per layer, GELU activations, layer normalization, and light dropout (0.013), producing 512-dimensional embeddings. The similarity between a household and a housing unit is computed using the dot product of their embeddings. The encoders are trained using a sigmoid-based contrastive loss \citep{zhai2023sigmoid}, which allows multiple valid matches per household. Training stability is enhanced using momentum distillation \citep{li2021align} with a housing-unit embedding queue (40,960 entries, decay = 0.83) and label smoothing (0.16). The contrastive loss $\mathcal{L}_{ij}^{\text{Contrastive}}$ is formulated as:
\begin{equation}
\mathcal{L}_{ij}^{\text{Contrastive}} = \log \frac{1}{1 + e^{-s(h_i, u_j)/\tau}}
\end{equation}
where $s(h_i, u_j)$ denotes the cosine similarity, and $\tau$ is a learnable temperature parameter that controls the concentration of the distribution.

In addition to contrastive learning, we incorporate an auxiliary instance-level Household-Unit Matching (HUM) decoder adapted from image-text matching models \citep{li2021align, radford2021learning}. This decoder takes concatenated household and housing embeddings $\{h_i, u_j\}$ and predicts a binary match probability using a four-layer feedforward network with residual connections. It is trained using positive pairs from true household-unit co-occurrences and hard-negative pairs sampled from within-cluster non-matches. The matching loss employs binary cross-entropy:
\begin{equation}
\mathcal{L}_{ij}^{\text{Match}} = y \log(s) + (1 - y) \log(1 - s),
\end{equation}
where $y \in \{0,1\}$ indicates ground-truth match status and $s$ denotes the decoder's output probability.

The matching loss is combined with the contrastive loss to balance global representation learning and local matching accuracy: $\mathcal{L} = \mathcal{L}^{\text{Contrastive}} + \lambda \mathcal{L}^{\text{match}}$, with $\lambda=0.5$ balancing global and local supervision.

The model is trained for 100 epochs using the AdamW optimizer \citep{loshchilov2017decoupled}, with a learning rate of $8.5 \times 10^{-5}$, momentum $\beta_1=0.999$, and weight decay $10^{-5}$. A cosine annealing schedule with a 5-epoch warmup is used to prevent early-stage training instability. The checkpoint achieving the best validation performance, measured by the Matthews correlation coefficient (MCC), is selected for inference because the validation set contains many more unobserved than observed housing-household pairs, and MCC remains informative under this class imbalance \citep{baldi2000assessing, chicco2020advantages}.

During inference, the trained encoders generate embeddings for all synthetic households derived from ACS PUMS data and all housing units from the NSI within each census tract. These embeddings are used to construct a dense compatibility matrix $\mathbf{S} \in \mathbb{R}^{H \times U}$, where $S_{hu} = \sigma(h \cdot u / \tau)$, representing the probability or compatibility that a given household $h$ can reside in a given housing unit $b$. Rather than applying hard thresholds, we retain the full probability distribution and pass it to a mixed-integer linear programming (MILP) allocation model that is introduced in Section \ref{sec:Household-to-Housing-allocation}.

\subsection{Household-to-Housing Unit Allocation Module}\label{sec:Household-to-Housing-allocation}

Each building contains one or more housing units. A household is assigned to a housing unit only if that unit is flagged as occupied; vacant units receive no assignment. The integration of synthetic population and building inventory is formulated as a hierarchical multi-objective mixed-integer linear program (MILP) that first generates an oversampled pool of synthetic households (50 $\times$ the target population size) and then selects a subset matching the ACS-derived population size and satisfying multiple criteria. Objectives are prioritized sequentially: (1) building-level feasibility (e.g., capacity constraints), (2) consistency with ACS marginal distributions, and (3) housing-household match quality, by maximizing aggregate compatibility scores across selected pairs.

\subsubsection{Rationale for Oversampling}\label{sec:oversampling_rationale}

The allocation step must select $n$ households from a candidate pool of size
$N$ that jointly satisfy building-capacity constraints, match CBG-level marginal targets, and maximize compatibility. Because the generator in Module 1 is trained on PUMS microdata, its output follows the PUMS joint distribution, which generally differs from CBG-level marginals. A pool of size $N = n$ would offer little combinatorial flexibility, leaving the optimizer few alternatives when a candidate satisfies one constraint but violates another. Oversampling at ratio $\alpha = N/n$ enlarges the integer program's feasible region by adding decision variables while keeping the constraint structure fixed.

Formally, let the candidate pool contain $X_k \sim \text{Binomial}(N, p_k)$ households in demographic category $k$, where $p_k$ is the category probability under the PUMS distribution. The target requires $n_k$ households in category $k$. By standard concentration inequalities \citep{mitzenmacher2005probability}, 
\begin{equation*}
    P(X_k < n_k) \leq \exp\!\bigl(-(Np_k - n_k)^2 / (2Np_k)\bigr), \text{whenever}~ Np_k > n_k,
\end{equation*}
and a union bound over all $K$ categories yields an overall infeasibility probability of at most $K \cdot \exp\!\bigl(-\Omega(\alpha \cdot p_{\min} \cdot n)\bigr)$, where $p_{\min}$ is the probability of the rarest category. Even moderate oversampling ratios ($\alpha = 5$-$10$) drive this probability to negligible levels, ensuring that the integer program's feasible region contains subsets matching the target distribution with high probability. The 50$\times$ ratio used here was empirically calibrated to guarantee feasibility across all CBGs in the study area, including those with rare demographic compositions.

\subsubsection{Sets and Indices}
\noindent We define the following sets that represent the core entities in our household-building allocation problem:

\medskip
\noindent\begin{tabularx}{\linewidth}{c X}
    \( H \) & Set of synthetic households, indexed by household ID (\texttt{HHID}). Each element represents a synthetic household with realistic sociodemographic characteristics and household composition derived from ACS microdata.\\
     \( B \) & Set of residential buildings from the NSI, indexed by building ID (\texttt{fd\_id}), each with known capacity and population constraints (e.g., presence of elderly individuals).\\
     \( \mathcal{A}_{\text{hh}} \) & Set of household-level attributes that characterize family units, including tenure status (\texttt{TEN}), vehicle availability (\texttt{VEH}), household income (\texttt{HINCP}), and other socioeconomic indicators that influence housing preferences.\\
     \( \mathcal{A}_{\text{p}} \) & Set of person-level attributes that describe each individual within a household, such as sex (\texttt{SEX}), age (\texttt{AGEP}), educational attainment (\texttt{SCHL}), and employment status, which together define household composition and housing needs.\\
     \( \mathcal{A}_{\text{u}} \) & Set of housing unit attributes, primarily occupancy status (occupied vs. vacant).
     \( C_a \): Set of all possible categories for attribute \( a \). For example, \( \texttt{TEN} \in \{1,2\} \) represents owner-occupied vs. renter-occupied housing, while \( \texttt{SEX} \in \{1,2\} \) represents male vs. female.
\end{tabularx}

\subsubsection{Parameters}
\noindent The model uses the following input parameters to represent key household attributes, building capacities, and target demographic distributions:

\medskip
\noindent\begin{tabularx}{\linewidth}{c X}
    \( s_{hb} \) &  Compatibility score between household \( h \in H \) and building \( b \in B \), predicted by the deep contrastive learning model \cite{qian2025deep}. Values range from 0 to 1, with higher scores indicating stronger household-building alignment.\\
    \( \text{NP}_h \) &  Household size, i.e., the number of individuals in household \( h \), relevant for space requirements and capacity constraints.\\
    \( \text{R65}_h \) &  Number of elderly individuals (age 65+) in household \( h \), capturing age-related housing selection.\\
    \( T^{\text{NP}}_b \) &  Target total population assigned to building \( b \), based on census block-level estimates.\\ 
    \( T^{\text{R65}}_b \) &  Target elderly population for building \( b \), ensuring alignment with ACS demographic distributions and NSI requirements.\\  
    \( U_b \) &  Maximum number of housing units in building \( b \), determined by NSI structural characteristics. \\
    \( \alpha_{a,c} \) &  Target proportion of household attribute \( a \in \mathcal{A}_{\text{hh}} \) in category \( c \in C_a \), derived from tract-level ACS data (e.g., \(\alpha_{\text{TEN},1}=0.65\) indicates 65\% owner-occupied households).\\ 
    \( \beta_{a,c} \) &  Target proportion of person attribute \( a \in \mathcal{A}_{\text{p}} \) in category \( c \in C_a \), representing individual-level demographic distributions (e.g., \(\beta_{\text{SEX},1}=0.49\) indicates 49\% male).\\ 
    \( \gamma_c \) &  Target proportion of housing unit occupancy status \( c \in \{0,1\} \), where \(1\)=occupied and \(0\)=vacant, reflecting local housing  stock. \\ 
    \( U_{\text{tot}} \) &  Total number of housing units across all buildings, including both NSI inventory and supplemental units not captured therein. 
\end{tabularx}

\subsubsection{Decision Variables}
\noindent The model defines the following decision variables:

\medskip
\noindent\begin{tabularx}{\linewidth}{c X}
    \( x_{hb} \in \{0,1\} \) &  Binary variable indicating whether household \( h \) is assigned to building \( b \) ($x_{hb}=1$) not not ($x_{hb}=0$). This indicator shows the allocation decision of mapping specific households to buildings. For example, if household 123 is placed in building 456, then \( x_{123,456} = 1 \) and \( x_{123,j} = 0 \) for all other buildings \( j \neq 456 \).\\
    \( y_h \in \{0,1\} \) &  Binary variable indicating whether household \( h \) is included in the final joint inventory ($y_h=1$) or excluded ($y_h=0$). Since the synthetic household pool is intentionally larger than available capacity, this variable ensures only a feasible subset of households is ultimately assigned.
\end{tabularx}

\noindent Additionally, slack variables are introduced to relax soft constraints, allowing the model to find feasible solutions even when an exact housing-household pair cannot be achieved:

\medskip
\noindent\begin{tabularx}{\linewidth}{c X}
    \( \zeta^{\text{NP}}_b, \zeta^{\text{R65}}_b \geq 0 \) &   Deviations from building-level target populations. For example, if building \( b \) should house 100 residents but only 95 are allocated, then \( \zeta^{\text{NP}}_b = 5 \). \\ 
    \( \varepsilon^{\text{hh}+}_{a,c}, \varepsilon^{\text{hh}-}_{a,c} \geq 0 \) &   Over- and under-representation of household attribute distributions. For instance, if the target share of owner-occupied households is 60\% but the allocation yields 65\%, then \(\varepsilon^{\text{hh}+}_{\text{TEN},1} = 0.05 \).\\ 
    \( \varepsilon^{\text{p}+}_{a,c}, \varepsilon^{\text{p}-}_{a,c} \geq 0 \) &   Deviations in person-level demographic distributions (e.g., sex, age, or education) across the allocated population.\\
    \( \varepsilon^{\text{u}+}_c, \varepsilon^{\text{u}-}_c \geq 0 \) &   Discrepancies between target and realized unit occupancy distributions, including vacancy rates.
\end{tabularx}

\subsubsection{Hierarchical Objective Functions}
The model adopts a three-level hierarchical objective structure, implemented through Gurobi's multi-objective optimization framework \citep{gurobi_2024}. Each stage is solved as a mixed-integer linear program via branch-and-cut, which combines LP-relaxation bounding, cutting-plane generation, and primal heuristics within a systematic search tree \citep{wolsey1998integer}. At termination, the solver returns a dual bound certifying that the solution lies within a specified optimality gap (MIPGap) of the global optimum, a guarantee that distinguishes the MILP formulation from heuristic or learning-based allocators, which converge to local optima without optimality certificates \citep{dauphin2014identifying, bengio2021machine}. The hierarchy prioritizes critical feasibility requirements first, then refines distributional accuracy, and finally maximizes the total matching score while preserving previously satisfied objectives. Slack variables provide controlled flexibility, ensuring feasible and realistic assignment of synthetic households to housing units across buildings.

\textbf{Objective 1 (Highest): Building capacity feasibility}
\begin{equation}
    \min \quad Z_1 = \sum_{b \in B} \left( \zeta^{\text{NP}}_b + \zeta^{\text{R65}}_b \right)
\end{equation}
The primary objective focuses on satisfying building capacity constraints in terms of both the total and the elderly population as closely as possible. Therefore, we aim to minimize deviations, $\zeta^{\text{NP}}_b$ and $\zeta^{\text{R65}}_b$, between the allocated synthetic population total and building capacity. A value of \( Z_1 = 0 \) indicates perfect satisfaction of all building-level requirements.

\textbf{Objective 2 (Secondary): Sociodemographics marginal distribution alignment}
\begin{equation}
\begin{aligned}
\min \quad Z_2 = & \sum_{a \in \mathcal{A}_{\text{hh}}, c \in C_a} \left( \varepsilon^{\text{hh}+}_{a,c} + \varepsilon^{\text{hh}-}_{a,c} \right) \\
& + \sum_{a \in \mathcal{A}_{\text{p}}, c \in C_a} \left( \varepsilon^{\text{p}+}_{a,c} + \varepsilon^{\text{p}-}_{a,c} \right) \\
& + \sum_{c \in \{0,1\}} \left( \varepsilon^{\text{u}+}_c + \varepsilon^{\text{u}-}_c \right)
\end{aligned}
\end{equation}
The secondary objective aims to minimize deviations from target demographic and socioeconomic distributions across three dimensions: household characteristics (e.g., tenure status, household income), individual characteristics (e.g., age, sex), and housing unit characteristics (e.g., occupancy status). This approach preserves the statistical consistency of the joined synthetic population with the ACS population marginal distributions.

\textbf{Objective 3 (Tertiary): Household and housing unit compatibility}
\begin{equation}
\min \quad Z_3 = -\sum_{h \in H} \sum_{b \in B} s_{hb} \cdot x_{hb}
\end{equation}
The tertiary objective maximizes the overall compatibility between the allocated synthetic households and their joined housing units. The objective is expressed with a negative sign to align with Gurobi's minimization-based multi-objective framework. It utilizes our proposed deep learning model's predicted compatibility scores \cite{qian2025dcl} to achieve the best possible housing-household pairings while adhering to the above two objectives' constraints.

\subsubsection{Constraints}

\textbf{Household and Housing Unit Consistency:}
\[
y_h = \sum_{b \in B} x_{hb}, \quad \forall h \in H
\]
This constraint ensures consistency between the allocation indicator \( x_{hb} \) and selection indicator \( y_h \). A household can only be assigned to a building ($x_{hb} =1$) if it is selected (from the oversampled pool) for inclusion in the final inventory ($y_h = 1$); if a household is not selected, it cannot be assigned anywhere. This constraint guarantees that only selected households are allocated, preventing mismatches where a household appears in the dataset without a valid assignment.

\textbf{Building-Level Constraints (with relaxation):}
\begin{equation}
\begin{aligned}
\sum_{h \in H} \text{NP}_h \cdot x_{hb} + \zeta^{\text{NP}}_b &= T^{\text{NP}}_b, \quad \forall b \in B \\
\sum_{h \in H} \text{R65}_h \cdot x_{hb} + \zeta^{\text{R65}}_b &= T^{\text{R65}}_b, \quad \forall b \in B \\
\sum_{h \in H} x_{hb} &\leq U_b, \quad \forall b \in B
\end{aligned}
\end{equation}
This constraint ensures that the total synthetic population allocated to building \( b \) equals the ACS ground-truth population size, with slack variable \( \zeta^{\text{NP}}_b \) permitting small deviations. The second constraint similarly regulates the elderly population distribution. The third constraint imposes a strict capacity limit, ensuring that the number of households assigned to a building does not exceed its maximum number of housing units. Unlike the total population constraints, this capacity condition cannot be violated, as it reflects a physical limitation.

\textbf{Household Attribute Distribution Constraints:}
Let \( N = \sum_{h \in H} y_h \) be the total number of selected households. Then:
\begin{equation}
\sum_{h \in H : \text{attr}_h(a) = c} y_h + \varepsilon^{\text{hh}-}_{a,c} - \varepsilon^{\text{hh}+}_{a,c} = N \cdot \alpha_{a,c}, \quad \forall a \in \mathcal{A}_{\text{hh}}, c \in C_a
\end{equation}
This constraint preserves the target distribution of household-level attributes within a selected population. For example, if the target proportion of households who owns the housing unit is \( \alpha_{\text{TEN},1} = 0.6 \), and there are 1000 households (\( N = 1000 \)) in total, then 600 should have owner as their tenure status, with allowable deviations represented by the slack variables \( \varepsilon^{\text{hh}+}_{\text{TEN},1} \) and \( \varepsilon^{\text{hh}-}_{\text{TEN},1} \).

\textbf{Person Attribute Distribution Constraints:}
Let \( P = \sum_{h \in H} \text{NP}_h \cdot y_h \) be the total selected population. Then:
\begin{equation}
\sum_{h \in H} \text{count}_h(a = c) \cdot y_h + \varepsilon^{\text{p}-}_{a,c} - \varepsilon^{\text{p}+}_{a,c} = P \cdot \beta_{a,c}, \quad \forall a \in \mathcal{A}_{\text{p}}, c \in C_a
\end{equation}
where \( \text{count}_h(a = c) \) represents the number of individuals in household \( h \) whose attribute \( a \) falls into category \( c \). This constraint ensures that individual-level characteristics (e.g., age, sex, education) are accurately represented in the allocated population. For example, if the target proportion of males, according to the ACS marginal distribution, is \( \beta_{\text{SEX},1} = 0.49 \) and the total allocated synthetic population is 5,000 individuals, then roughly 2,450 males should appear across all households in the final joint inventory.

\textbf{Unit Occupancy Distribution Constraints:}
\begin{equation}
\begin{aligned}
N + \varepsilon^{\text{u}-}_1 - \varepsilon^{\text{u}+}_1 &= U_{\text{tot}} \cdot \gamma_1 \\
(U_{\text{tot}} - N) + \varepsilon^{\text{u}-}_0 - \varepsilon^{\text{u}+}_0 &= U_{\text{tot}} \cdot \gamma_0
\end{aligned}
\end{equation}
These constraints regulate the overall vacancy rate of the housing units. The first ensures that the number of occupied units (\( N \), as each household occupies one unit) aligns with the target occupancy rate \( \gamma_1 \). The second that the number of vacant units (\( U_{\text{tot}} - N \)) aligns with the target vacancy rate \( \gamma_0 \). Together, they maintain a realistic housing stock where \( \gamma_1 + \gamma_0 = 1 \). The target rates \( \gamma_0 \) and \( \gamma_1 \) are derived from ACS estimates of housing unit vacancy at the census block group level, ensuring that the optimized inventory reproduces locally observed vacancy patterns.

\section{Experiment Configurations}

\subsection{Data Wrangling for PUMS and NSI Variable Consistency}
\label{sec:data_prep}

After learning housing-household relationships from ACS PUMS microdata, we apply them to link the synthetic population to a real housing inventory. This integration is complicated because ACS PUMS and the NSI describe related concepts using different variables, definitions, and spatial resolutions. ACS PUMS reports sociodemographic and selected housing-unit attributes at the person and household level, whereas the NSI describes physical structures at the building level. A single-family home typically corresponds to one housing unit, but multi-family buildings contain several, so NSI records cannot be treated directly as housing units. The two sources also use inconsistent variable definitions (e.g., property value in PUMS versus structure value in NSI).

To reconcile these differences, we align housing-related variables across ACS PUMS, the synthetic population, and the NSI so they are conceptually and statistically comparable. The preprocessing pipeline involves two steps: disaggregating NSI buildings into individual housing units and harmonizing core housing attributes to ensure compatibility with the learned housing-household relationships.

\subsubsection{Property Value (VALP) Alignment}

The NSI reports a total structural value for each building (\texttt{val\_struct}), whereas our joining framework requires housing unit-level property values (\texttt{VALP}). To derive an initial unit-level estimate, we divide the total building value by the estimated number of housing units within the structure (\texttt{num\_units}): 
\begin{equation}
    \text{VALP}_{\text{est}} = \frac{\text{val\_struct}}{\text{num\_units}}
\end{equation}

This preliminary estimate may be biased because NSI valuation methods differ systematically from the self-reported property values in the ACS. To correct for this discrepancy, we calibrate the estimated unit values so that their distribution aligns with the \texttt{VALP} distribution observed in the ACS PUMS, denoted by $g$. Specifically, we apply a linear transformation and estimate its parameters by matching the 5th and 95th percentiles of the two distributions:
\begin{equation}
\alpha = \frac{g_{95} - g_{05}}{v_{95} - v_{05} + 10^{-8}}, \quad \beta = g_{95} - \alpha v_{95}
\end{equation}
where $v_{05}$ and $v_{95}$ are the percentiles for $\text{VALP}_{\text{est}}$, and $g_{05}$ and $g_{95}$ are the corresponding percentiles for empirical \texttt{VALP} in the ACS PUMS. The calibrated housing-unit property value is then computed as 
\begin{equation}
\text{VALP}_{\text{cal}} = \alpha \cdot \text{VALP}_{\text{est}} + \beta
\end{equation}
To ensure realism and avoid implausibly low values, we cap the calibrated estimate at the minimum observed \texttt{VALP} in the ACS ($g_{\text{min}}$): 
\begin{equation}
\text{VALP} = max(\text{VALP}_{\text{cal}}, g_{\text{min}})
\end{equation}

Because the NSI records structural replacement cost rather than market transaction price or rent, VALP is the sole monetary attribute on the housing side of the matching model. Gross rent (\texttt{GRNTP}) is recorded in ACS PUMS at the housing-unit level, but the NSI contains no rent-related field, so we rely on property value as a monetary proxy for both owner- and renter-occupied units. To assess whether this proxy introduces systematic bias for renter households, we examined the empirical VALP--GRNTP relationship in ACS PUMS. After conditioning on the structural attributes used by the matching model, namely building type (\texttt{BLD}), bedroom count (\texttt{BDSP}), construction vintage (\texttt{YRBLT}), and geographic location (\texttt{PUMA}), the within-stratum Spearman rank correlation is $\rho = 0.807$ across 235 strata, and the Pearson correlation of the log-transformed values is $r = 0.750$. At the PUMA level, these correlations rise to $\rho = 0.870$ and $r = 0.878$. Hedonic regressions of $\log(\text{VALP})$ and $\log(\text{GRNTP})$ on these attributes yield consistent signs and comparable magnitudes for all variables except lot acreage (\texttt{ACR}), whose divergence (positive for owners, reflecting resale land value; negative for renters, reflecting peripheral location) is economically expected and contributes only marginally to match scores. At the housing unit level, VALP therefore preserves the ordinal ranking of housing units that GRNTP would produce, and the absence of an explicit rent variable does not materially distort compatibility estimates for renter households.

\subsubsection{Land Acreage (ACR) Estimation}

The ACS PUMS lot size variable (ACR) classifies housing units into four ordinal categories (e.g., 0: missing or multifamily, 1: $<$1 acre, 2: 1-9.99 acres), and 3: $\geq$ 10 acres), but the NSI does not report lot size directly. Rather than estimating exact acreage, we infer ACR categories using structural and contextual features from the NSI, while constraining the aggregate category distribution to match that observed in ACS. 

We derive five normalized factors from NSI variables: building square footage (\texttt{sqft}), occupancy type (\texttt{occtype}), structural value (\texttt{val\_struct}), building material (\texttt{bldgtype}), and construction year (\texttt{med\_yr\_blt}). These factors capture systematic relationships between building characteristics, development patterns, and parcel size and are combined to infer ordinal lot-size categories.

These factors are combined into a composite score $S$ using fixed weights informed by engineering judgment and data-driven considerations:
\begin{equation}
S = 0.35 f_{\text{sqft}} + 0.25 f_{\text{type}} + 0.20 f_{\text{value}} + 0.10 f_{\text{bldg}} + 0.10 f_{\text{year}}
\end{equation}

Next, we derive the empirical distribution of acreage categories from the ACS microdata, computing the proportions $(p_1, p_2, p_3)$ or ACR categories 1, 2, and 3, respectively. All housing units are then ranked by their composite score $S$, and threshold values $\tau_1$ and $\tau_2$ are determined such that the cumulative shares of units match the ACS proportions $p_1$ and $p_1 + p_2$. 

Each housing unit is assigned an acreage category according to:
\begin{equation}
\text{ACR} =
\begin{cases}
1, & S < \tau_1; \\
2, & \tau_1 \leq S < \tau_2; \\
3, & S \geq \tau_2; \\
0, & \texttt{occtype} = \texttt{RES2}.
\end{cases}
\end{equation}
Manufactured housing units (\texttt{RES2}) are directly assigned ACR = 0, as they typically do not correspond to privately owned lots in typical cases. 

\subsubsection{Building Structure Type (BLD) Estimation}

The ACS building structure variable (BLD) categorizes structures by the number of housing units (1-9), from mobile homes to large apartment buildings. This information is not directly available in the NSI, which instead reports occupancy type (\texttt{occtype}) and estimated unit counts (\texttt{num\_units}). We retain only residential structures: single-family homes, mobile homes, and multi-family buildings with 2-50 units, and exclude non-residential or group-quarter types (e.g., hotels, dormitories, nursing homes). ACS-consistent BLD categories are inferred from NSI variables using a rule-based, data-informed mapping. Our inference algorithm proceeds as follows:

\begin{enumerate}
    \item \textbf{Mobile homes} (BLD = 1). Structures classified as manufactured houses (\texttt{occtype} = RES2) are assigned BLD = 1. This direct mapping reflects the distinct definition of mobile homes in the ACS.
    \item \textbf{Single-family homes} (BLD = 2 or 3). For single-family homes (\texttt{occtype} = RES1), we assign detached (BLD = 2) or attached (BLD = 3) categories by sampling from the ACS PUMS distribution, where roughly 82\% are detached and 18\% attached, since NSI does not distinguish these subtypes.
    \item \textbf{Multi-family housing} (BLD = 4-9). For multi-family residential buildings, BLD categories are inferred based on the estimated number of units (\texttt{num\_units}), following ACS definitions:
    \begin{itemize}
        \item 2 units: BLD = 4
        \item 3-4 units: BLD = 5
        \item 5-9 units: BLD = 6
        \item 10-19 units: BLD = 7
        \item 20-49 units: BLD = 8
        \item 50 or more units: BLD = 9
    \end{itemize}
    When a building's unit count spans a range that maps to multiple BLD categories, we sample the final assignment using conditional probabilities $Pr(\text{BLD} \mid \text{num}_\text{units})$ estimated from ACS microdata. This preserves realistic structure-type distributions within each unit-count class.
    \item \textbf{Fallback assignment}. In rare cases where occupancy type or unit count information is insufficient or inconsistent, we assign BLD by sampling from the overall marginal distribution of BLD observed in the ACS. This ensures that every housing unit receives a valid structure type and avoids missing values in the inventory.
\end{enumerate}

Together, this hybrid rule-based and probabilistic approach produces BLD assignments that are fully compatible with ACS definitions while remaining grounded in the structural information available in the NSI.

\subsubsection{Disaggregation of Household and Person Attributes}

NSI population data are reported at the building level and require different treatment for single- and multi-unit structures. For each building, we preserve key NSI constraints (total population and the number of residents aged 65 and over) and disaggregate them to the housing-unit level.

For single-family homes, NSI population counts are assigned directly to unit-level variables. For multi-family buildings, where NSI provides only aggregate counts, we apply a probabilistic disaggregation informed by ACS microdata: unit-level household size is first sampled from the ACS distribution, and elderly residents are then assigned using the conditional distribution $Pr(\text{R65}\mid\text{NP})$. Rare combinations are handled by extrapolation, and sampled assignments are adjusted to maintain consistency with NSI-reported building totals. Disability counts by age group are subsequently generated at the unit level via binomial sampling based on NSI disability rates.

During disaggregation, some units receive a simulated occupant count of zero. In single-family buildings with zero NSI population, the unit is directly flagged as vacant. In multi-family buildings, probabilistic sampling from ACS household-size distributions (which include zero-person outcomes reflecting unoccupied units) may assign zero occupants to individual units even when the building-level total is positive. These preliminary counts serve two downstream purposes: they provide the matching model with a unit-level occupancy feature (NP), and they allow the candidate-filtering step (Section~\ref{sec:Household-to-Housing-allocation}) to exclude clearly vacant units from assignment. The simulated NP values are provisional; the optimization step determines actual occupancy through its household-assignment decisions. The optimizer may therefore leave units vacant that received a positive simulated NP, and the realized vacancy pattern is ultimately governed by the unit-occupancy distribution constraints (Section~\ref{sec:Household-to-Housing-allocation}) rather than by the preprocessing estimates alone.

\subsubsection{Construction Year (YRBLT) Categorization}

The ACS represents the year of construction using eight ordinal categories (\texttt{YRBLT} = 0-7), each corresponding to a historical building cohort associated with distinct structural standards and vulnerability characteristics. We map the NSI variable \texttt{med\_yr\_blt} to these categories using deterministic thresholds aligned with ACS definitions as below. For missing or implausible values, we impute the modal \texttt{YRBLT} category within the same census tract to maintain spatial consistency.

\begin{equation}
    \text{YRBLT} =
    \begin{cases}
        0, & \texttt{year} \le 1940;\\
        1, & 1941 \le \texttt{year} \le 1950;\\
        2, & 1951 \le \texttt{year} \le 1960;\\
        3, & 1961 \le \texttt{year} \le 1970;\\
        4, & 1971 \le \texttt{year} \le 1980;\\
        5, & 1981 \le \texttt{year} \le 1990;\\
        6, & 1991 \le \texttt{year} \le 2010;\\
        7, & \texttt{year} > 2010.
    \end{cases}
\end{equation}

\subsubsection{Output and Data Export}

Upon completion of the above preprocessing steps, we export a housing unit-level NSI table containing \texttt{VALP,ACR,NP,R65,o65DIS,u65DIS,YRBLT,} and \texttt{BLD}, along with the geo-identifiers required by the allocation model. The table is stored in state-specific directories in Parquet format to ensure both storage efficiency and fast downstream access. This dataset provides a statistically aligned, attribute-complete housing unit inventory that supports the household-to-unit joining framework described in Section~\ref{sec:Household-to-Housing-allocation}.

\subsection{Search Space Reduction and Optimization Setup}

We generate a candidate household pool roughly 50 times larger than the required population for assignment. Directly solving a full bipartite matching problem between households ($\mathcal{H}$) and housing units ($\mathcal{U}$) is computationally infeasible, so we apply a two-step filtering strategy for each unit $u \in \mathcal{U}$.

First, households with matching scores $s(u,h)$ below a threshold $\tau$ are discarded. By default, $\tau = 0.5$, relaxed to 0.2 in block groups requiring large allocations (e.g., $>$900 households) to maintain a sufficient candidate pool. Second, only the top-$K$ households by score are retained, with $K$ set adaptively based on the number of households to allocate ($\mathcal{H}_{\text{req}}$): 8,000 for $\mathcal{H}_{\text{req}} < 800$, increasing to 32,000 for $\mathcal{H}_{\text{req}} > 2,600$. Candidates are aggregated at the building level to enforce population constraints, and only these ``high-qualit'' pairs define binary decision variables in the optimization model. All thresholds and parameters were empirically determined from pilot experiments.

The optimization in the housing-to-unit allocation module (Sec. \ref{sec:Household-to-Housing-allocation}) is solved using Gurobi with stage-specific parameter settings designed to balance feasibility, solution quality, and runtime. 

\begin{itemize}
    \item \textit{Stage 1 (Building capacity feasibility):} This stage focuses on quickly identifying a feasible assignment that satisfies all building-level constraints. We impose an 8-hour running time limit (28,800 sec), set \texttt{MIPFocus} to 1 to emphasize feasibility, and use a \texttt{MIPGap} of 0.009. To accelerate the search, we enable aggressive heuristics (\texttt{Heuristics=0.5}, \texttt{PumpPasses=30}, \texttt{RINS=30}).
    \item \textit{Stage 2 (Population marginal distribution alignment):} Once feasibility is achieved, the second stage targets alignment between the marginal distributions of household attributes and census benchmarks. This stage is allotted a 12-hour time limit (43,200 sec) and retains the same \texttt{MIPFocus} and \texttt{MIPGap} settings as Stage 1 to maintain solution quality.
    \item \textit{Stage 3 (Household and housing unit compatibility maximization):} The final stage refines the assignment by maximizing the aggregate matching score. We impose a 5-hour time limit (18{,}000 sec) and adopt a more relaxed solver configuration, with \texttt{MIPFocus=0} and a larger \texttt{MIPGap} of 0.05, to allow broader exploration of the solution space.
\end{itemize}

Each three-stage optimization is defined for a single census block group, so the possible 25-hour total represents a worst-case cap per block-group instance rather than for an entire tract or the full nine-county region. Regional production repeats this solve across many block groups, with total wall-clock time depending on the block-group size distribution and the available parallel computing resources.

\section{Results}

Using the proposed framework, we construct a joint housing-household inventory for 9 coastal counties in North Carolina \cite{nc_eastern_zone}. The performance of each module is evaluated individually to ensure methodological transparency and to facilitate the adoption of both the framework and the resulting joint inventory in disaster impact assessment and community resilience planning.

\subsection{Synthetic Population Generation Performance}

The fidelity of synthetic household data directly affects the quality of downstream allocation. To evaluate generation performance, we use metrics from the sdmetrics library \citep{sdmetrics}, comparing our method, TabDiff, against two widely adopted baselines from the Synthetic Data Vault (SDV) framework: TVAE (Tabular Variational Autoencoder with conditional sampling) and CTGAN (Conditional GAN with mode-specific normalization) \citep{xu2019modeling, sdmetrics}. These baselines are standard benchmarks for mixed-type tabular synthesis. The evaluation metrics and comparative results are summarized below.

\begin{itemize}
    \item \textit{\textbf{Column Shape Similarity (Shape):}} Measures alignment in univariate distributions. For numerical attributes, the score is $1 - D_{KS}$, where $D_{KS} = \max_{x} |F_{real}(x) - F_{synth}(x)|$ is the Kolmogorov-Smirnov statistic between empirical cumulative distribution functions (CDFs). For categorical columns, the score is $1 - \text{TVD}$, where $\text{TVD} = \frac{1}{2} \sum_{k} |p_{real,k} - p_{synth,k}|$ is the Total Variation Distance (TVD) over category probabilities.
    
    \item \textit{\textbf{Column Pair Trend Similarity (Trend):}} Evaluates preservation of bivariate relationships. For numerical pairs, the score is $1 - \frac{|\rho_{real} - \rho_{synth}|}{2}$, where $\rho$ is the Pearson correlation. For categorical pairs, the score is $1 - \text{TVD}_{contingency}$, where $\text{TVD}_{contingency} = \frac{1}{2} \sum_{i,j} |P_{real}(cat_i, cat_j) - P_{synth}(cat_i, cat_j)|$ is the TVD over joint distributions.
    
    \item \textit{\textbf{Overall Quality (Overall)}}: The arithmetic mean of Shape and Trend, providing a single composite fidelity metric.
    
    \item \textit{\textbf{Classifier Two-Sample Test (C2ST):}} Measures indistinguishability by training a logistic regression classifier to separate real samples from synthetic data. The score is $1 - |2 \times \text{AUC} - 1|$, where AUC is the area under the ROC curve. A value of 1 corresponds to AUC $= 0.5$, indicating random guessing and ideal indistinguishability, while 0 indicates perfect separation.
    
    \item \textit{\textbf{$\alpha$-Precision:}} measures the proportion of synthetic households that remain within the plausible region of the real population data (i.e., PUMS) distribution. Intuitively, if a generative model produces households with unrealistic demographics (e.g., a 18-year-old listed as the householder with a doctoral degree and \$500{,}000 income), such records cannot represent a realistic synthetic population. These out-of-distribution samples would propagate errors into the housing-household matching step. We therefore require a metric that flags whether synthetic population remain within the plausible region of the real data distribution. Following the sdmetrics library \citep{sdmetrics}, $\alpha$-Precision estimates this proportion using $k$-nearest neighbor density estimation: a synthetic sample is considered plausible if its distance to the nearest real sample falls below a threshold $\alpha$ that delineates the high-density region of the true distribution.
    Formally,
    \begin{equation}
        \alpha\text{-Precision} = \frac{|\{\mathbf{x} \in G : \min_{\mathbf{y} \in R} d(\mathbf{x},\mathbf{y}) < \alpha\}|}{|G|}
    \end{equation}
    where $G$ is the set of synthetic samples, $R$ is the set of real microdata samples, $d(\mathbf{x},\mathbf{y})$ is the distance metric in feature space, and $\alpha$ specifies the realism threshold defining what qualifies as a realistic sample. A high $\alpha$-Precision indicates that the generator rarely produces implausible or out-of-support households, an essential requirement for constructing coherent joint housing-household inventories. 
    
    \item \textit{\textbf{$\beta$-Recall:}} complements precision by assessing the diversity of the synthetic population. It quantifies the proportion of real samples that lie within the $beta$-support of the synthetic distribution, i.e., how well the model captures the full variability observed in the true population:
    \begin{equation}
        \beta\text{-Recall} = \frac{|\{\mathbf{y} \in R : \min_{\mathbf{x} \in G} d(\mathbf{x},\mathbf{y}) < \beta\}|}{|R|}
    \end{equation}
    where $\beta$ sets the inclusion threshold. High $\beta$-Recall indicates good coverage of population heterogeneity and resistance to mode collapse (i.e., the generator only learns the most common modes and fails to capture the true diversity). 
\end{itemize}

\paragraph{Implementation.}

We implement TVAE and CTGAN using the SDV library \citep{sdv}. TVAE uses a variational autoencoder with two 128-dimensional hidden layers for encoder and decoder, trained for 300 epochs with batch size 500 and L2 regularization ($\lambda=10^{-5}$). CTGAN employs a conditional GAN with two 256-dimensional hidden layers for generator and discriminator, trained for 300 epochs with learning rates $2\times10^{-4}$ and PACGan augmentation (PAC size 10). Both models enforce min-max constraints and rounding rules to preserve data types. For each method, we train 10 models with different seeds and generate 44,495 synthetic samples per run, matching the training set size for robust comparisons.

\paragraph{Performance Evaluation}
Table \ref{tab:generative_model_performance} 
shows the performance of the three generative models over 10 independent runs. 

\begin{table}[!h]
\centering
\caption{Comparative Evaluation of Synthetic Household Data Generation Methods.}
\label{tab:generative_model_performance}
\resizebox{\textwidth}{!}{%
\begin{tabular}{@{}lcccccc@{}}
\toprule
Model & Shape $\uparrow$ & Trend $\uparrow$ & Overall $\uparrow$ & C2ST $\uparrow$ & Alpha Prec. $\uparrow$ & Beta Recall $\uparrow$ \\
\midrule
TabDiff & $\mathbf{0.9982 \pm 0.0001}$ & $\mathbf{0.9417 \pm 0.0283}$ & $\mathbf{0.9699 \pm 0.0141}$ & $\mathbf{0.9890 \pm 0.0037}$ & $\mathbf{0.9934 \pm 0.0017}$ & $\mathbf{0.4798 \pm 0.0028}$ \\
TVAE    & $0.9690 \pm 0.0028$ & $0.7184 \pm 0.0047$ & $0.8437 \pm 0.0036$ & $0.5161 \pm 0.0339$ & $0.9177 \pm 0.0181$ & $0.1646 \pm 0.0138$ \\
CTGAN   & $0.9127 \pm 0.0048$ & $0.7710 \pm 0.0086$ & $0.8418 \pm 0.0066$ & $0.5235 \pm 0.0788$ & $0.7620 \pm 0.0505$ & $0.1736 \pm 0.0212$ \\
\bottomrule
\end{tabular}
}
\end{table}

\textbf{\textit{Univariate and bivariate realism.}} TabDiff achieves excellent alignment with the real microdata when looking at each variable's marginal distribution (Shape = $0.9982 \pm 0.0001$). More importantly, it preserves the relationships between pairs of variables (Trend = $0.9417 \pm 0.0283$), which is critical for maintaining realistic correlations among household-level and individual-level attributes. TVAE performs reasonably well on univariate distributions (Shape = $0.9690 \pm 0.0028$) but fails to capture pairwise dependencies (Trend $= 0.7184 \pm 0.0047$). CTGAN performs even worse on both fronts (Shape = $0.9127\pm0.0048$, Trend = $0.7710\pm0.0086$), indicating substantial distortions in both single-variable and multi-variable relationships.

\textbf{\textit{Indistinguishability from real data.}} The C2ST metric shows how easily a classifier can separate synthetic samples from real ones. TabDiff obtains the highest score ($0.9890 \pm 0.0037$), meaning that its generated synthetic data are very difficult to distinguish from the real population data. In contrast, both TVAE and CTGAN's C2ST score hover near 0.5, essentially signaling that their generated synthetic data can easily be flagged as fake.

\textbf{\textit{Precision-recall tradeoff (realism vs. diversity).}} TabDiff strikes a balance between producing realistic samples and covering the full range of real population patterns. Its high $\alpha$-Precision ($0.9934 \pm 0.0017$) shows that almost all of its synthetic households fall within the plausible region of real population distribution. At the same time, its $\beta$-Recall ($0.4798 \pm 0.0028$) indicates decent coverage of the population diversity. By contrast, both TVAE and CTGAN achieve very low recall ($0.17$), meaning that they fail to represent large portions of the real population distribution, a classic sign of mode collapse, where only the ``easy-to-mode'' portions of the population are generated while ignoring rarer but important cases.

Across every dimension of evaluation, such as realism, diversity, feature relationships, and statistical indistinguishability, TabDiff substantially outperforms both the VAE-based (TVAE) and GAN-based (CTGAN) baselines. This is important for constructing a realistic synthetic joint housing-household inventory, as it requires accurate correlations among household and individual attributes (e.g., income, education, age, and tenure) to be preserved. TabDiff's superior performance ensures these relationships are preserved, enabling more accurate and demographically consistent allocation of households to housing units.

\subsection{Housing-household Relationship Learning Performance}

\paragraph{Evaluation Metrics.} The matching module directly affects how well the downstream allocation meets demographic targets. We assess its performance using stratified confusion matrices (TP, TN, FP, FN), with particular emphasis on the Matthews correlation coefficient (MCC). MCC is the $\phi$ correlation between predicted and observed binary labels, incorporating all four confusion-matrix entries rather than a single error type \citep{baldi2000assessing, chicco2020advantages}. Following standard machine-learning practice, including the scikit-learn definition of MCC as a balanced measure for classes of unequal size, we interpret MCC on the interval [-1,1], where 1 indicates perfect agreement, 0 denotes chance-level prediction, and -1 shows systematic disagreement. This choice fits our setting because each observed housing-household pair is evaluated against a much larger set of unobserved pairs, and the positive match rate varies across tracts after candidate filtering. Under such an imbalance, accuracy can remain high even when minority-class performance is poor, and F1 captures precision and recall for the positive class but ignores true negatives. MCC decreases whenever errors accumulate in either class, providing a more complete summary of how well the model separates plausible from implausible matches. While MCC is less familiar in built-environment and engineering applications than accuracy or F1, it is a well-established metric for imbalanced classification in machine learning. Model performance is therefore evaluated using the following standard binary classification metrics:
\begin{itemize}
    \item \textbf{Accuracy}: The proportion of housing-household unit pairs that are correctly classified among all evaluated pairs. \vspace{-4pt}
    \item \textbf{Area Under the ROC Curve (AUC)}: The probability that the model assigns a higher score to a true matching pair than to a non-matching pair across all decision thresholds.\vspace{-4pt}
    \item \textbf{Precision}: For a given class, defined as $\text{TP} / (\text{TP} + \text{FP})$, measuring the reliability of predicted plausible matches.\vspace{-4pt}
    \item \textbf{Recall}: For a given class, defined as $\text{TP} / (\text{TP} + \text{FN})$, capturing the fraction of true plausible matches correctly identified by the model.\vspace{-4pt}
    \item \textbf{F1-Score}: The harmonic mean of precision and recall, balancing FP and FN.\vspace{-4pt}
    \item \textbf{Matthews Correlation Coefficient (MCC)}: A confusion-matrix correlation coefficient computed as $\frac{\text{TP} \times \text{TN} - \text{FP} \times \text{FN}}{\sqrt{(\text{TP} + \text{FP})(\text{TP} + \text{FN})(\text{TN} + \text{FP})(\text{TN} + \text{FN})}}$, which remains robust under class imbalance and varying match densities across census tracts. 
\end{itemize}

\paragraph{Housing-Household Matching Performance} Table~\ref{tab:matching_model_performance} shows that the matcher achieves high fidelity across all measures, providing the reliable compatibility scores required by the allocation optimization. The accuracy of 0.9267 and AUC of 0.9725 indicate strong discrimination, and the class-specific precision and recall values remain well balanced on both observed matches and unobserved pairs. The MCC of 0.8536 is particularly informative because it penalizes false positives and false negatives in both classes simultaneously. A value above 0.85 reflects strong agreement between predicted and observed labels rather than a favorable score driven by class imbalance. The model therefore learns a meaningful separation between plausible and implausible housing-household pairs, which is exactly the quantity required by the downstream allocation stage.

\begin{table}[htbp]
\centering
\caption{Performance Evaluation of the Housing-Household Relationship Learning}
\label{tab:matching_model_performance}
\begin{tabular}{llllllllll}
\hline
Metric & \multicolumn{1}{c}{Accuracy} & \multicolumn{1}{c}{AUC} & \multicolumn{2}{c}{F1-Score} & \multicolumn{2}{c}{Precision} & \multicolumn{2}{c}{Recall} & \multicolumn{1}{c}{MCC} \\ \cline{4-9}
       &                              &                         & Class 0       & Class 1      & Class 0       & Class 1       & Class 0      & Class 1     &                         \\ \hline
Value  & 0.9267                       & 0.9725                  & 0.9259        & 0.9275       & 0.9359        & 0.9179        & 0.9161       & 0.9373      & 0.8536                  \\ \hline
\end{tabular}
\end{table}

\subsection{Joint Housing-Household Inventory Performance}

\begin{figure}[htbp]
    \centering
    \includegraphics[width=0.95\textwidth]{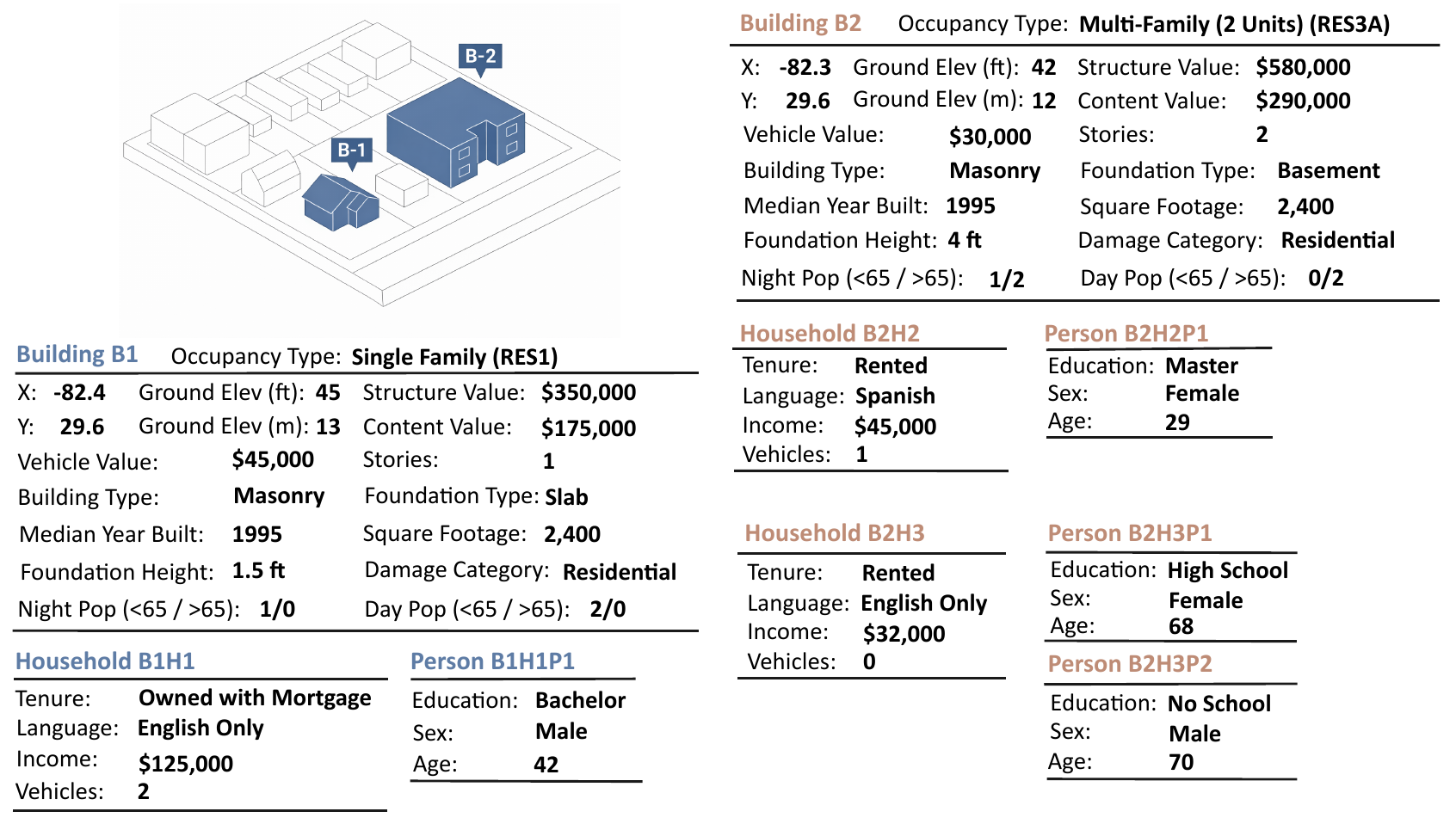}
    \caption{Illustrative example of the joint housing-household inventory.}
    \label{fig:inventory_structure}
\end{figure}

We develop a joint housing-household inventory for Eastern North Carolina that links detailed housing characteristics with household- and individual-level information by combining NSI structural attributes with a synthetic population generated from ACS PUMS data. Each housing unit is associated with one or more households, and each household is linked to individuals living within it, each carrying their own attributes. Figure~\ref{fig:inventory_structure} illustrates the inventory using a randomly selected CBG within Census Tract 37051001903, Cumberland County, North Carolina, including both single-family homes and multi-family buildings to demonstrate the framework's ability to represent diverse residential settings.

The validation evidence operates at three levels. First, the synthetic population module is benchmarked against real ACS PUMS microdata using distribution alignment, real vs. synthetic population indistinguishability, and support-coverage metrics. Second, the compatibility model is evaluated against held-out observed housing-household pairs from ACS PUMS, achieving 92.7\% accuracy and an MCC of 0.854. Third, the final allocation is compared with real external targets at the highest publicly available spatial resolution: NSI building-level population counts and ACS block-group-level marginal distributions of household, person, and occupancy variables. This design anchors each pipeline stage to observed data, while direct address-level housing-household linkage validation remains outside the scope of the present study.

In this example, building \textit{B1} is a one-story masonry single-family home with 2,400 square feet, occupied by a one-person household. The household has an annual income of about \$125,000, owns the home (built around 1995 on a 1.5-ft slab foundation, with a structural value of \$350,000), speaks English only, and owns two vehicles. The resident is a 42-year-old male with a Bachelor's degree. 

Building \textit{B2} is a two-story, two-unit masonry structure built around 1995, with a 4-ft foundation with basement, 2,400 square feet, and a structural value of \$580,000. It houses two households. The first is a 29-year-old female renter earning approximately \$34,000, primarily Spanish-speaking, with one vehicle and a master's degree. The second household earns approximately \$32,000, rents the unit, speaks English only, owns no vehicle, and consists of two elderly adults (a 70-year-old male with no formal education and a 68-year-old female with a high school education).

This within-building contrast highlights the socioeconomic diversity captured by the joint inventory: residents sharing the same structure may have very different needs, resources, and capacities to prepare for, respond to, and recover from disasters. By preserving these detailed household-structure relationships, the inventory supports more targeted, equitable social assistance and resilient planning interventions.

\subsubsection{Population Density Match}

\begin{figure}[htbp]
    \centering
    \includegraphics[width=0.95\textwidth]{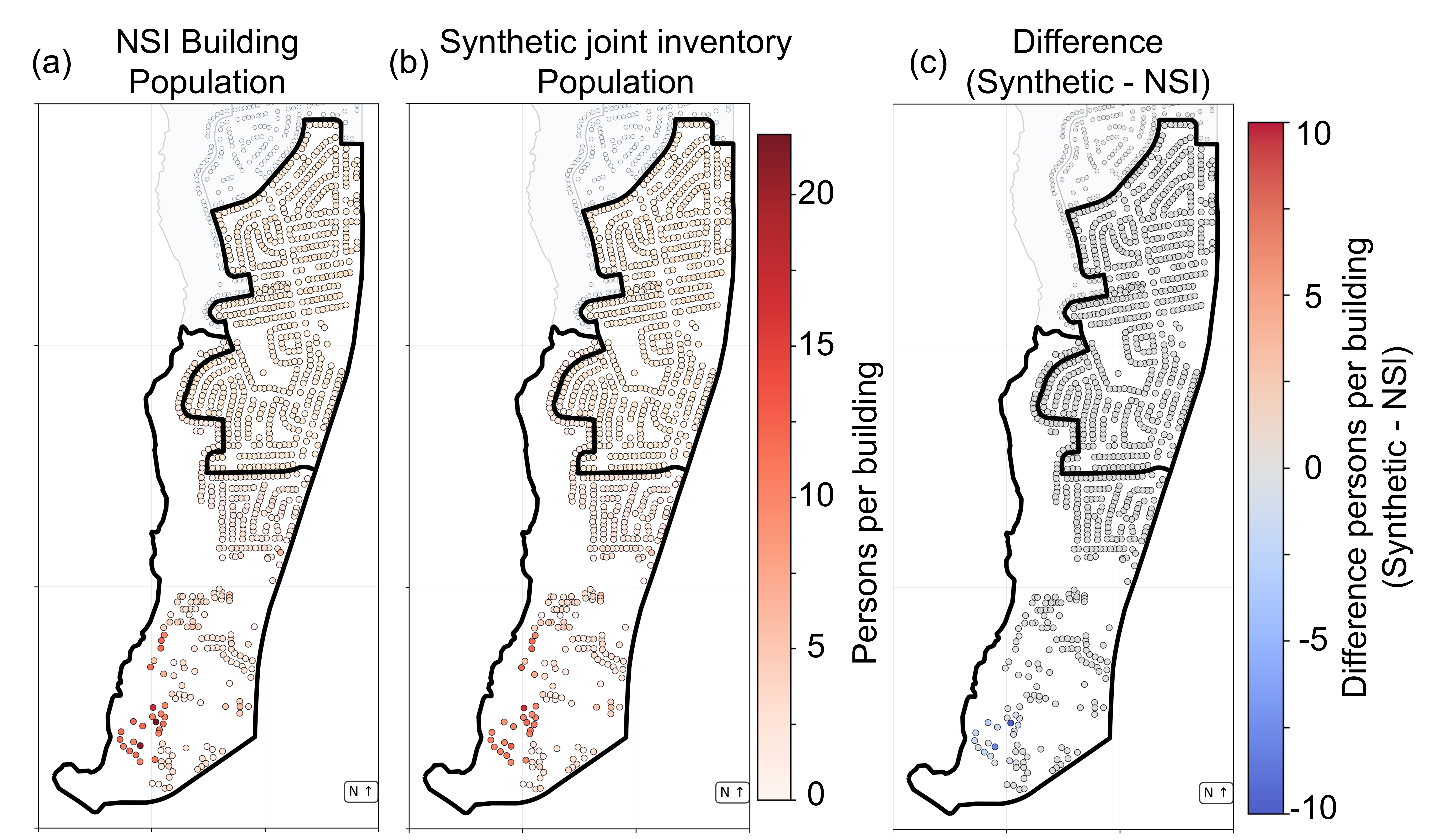}
    \caption{Spatial distribution of building-level population in census block groups 370510019032 and 370510019033. }
    \label{fig:spatial_population_distribution}
\end{figure}

We analyze the spatial patterns of the allocated synthetic population and the ground-truth NSI population at the building level within Census Tract 37051001903 in Cumberland County, North Carolina. This tract includes two census block groups (CBGs 370510019032 and 370510019033) and represents a typical suburban area with a mix of housing types. Together, these two CBGs contain 840 residential buildings. Building-level ground-truth population counts from the NSI are computed by summing the populations under age 65 (pop2pmu65) and over age 65 (pop2pmo65), yielding the expected total occupancy for each structure. Figure~\ref{fig:spatial_population_distribution} compares the NSI population distribution with the synthetic population produced by our framework.

Figure~\ref{fig:spatial_population_distribution}(a) shows the NSI building-level population, revealing a heterogeneous spatial pattern: single-family homes with one to two residents are concentrated in the southern portion of the tract, while buildings with higher occupancy (three to four residents) are more common in the northern area. Figure~\ref{fig:spatial_population_distribution}(b) presents the corresponding synthetic population allocation. The synthetic results closely replicate the overall spatial distribution observed in the NSI data, while also capturing realistic variation at the household level. In particular, the allocation preserves the clustering of similar household sizes, which emerges naturally from the learned compatibility between household characteristics and housing unit features.

Figure~\ref{fig:spatial_population_distribution}(c) shows the differences between the synthetic and NSI populations (Synthetic - NSI), where blue indicates fewer allocated residents and red indicates more allocated residents than the NSI benchmark. These differences are small: 89\% of buildings differ by no more than one person, and the average difference across the tract is $-0.03$ persons per building. Larger deviations occur mainly in multi-unit buildings, where disaggregating building-level populations into individual units introduces modest, controlled variation. However, these differences balance out at the CBG level, with total populations within 0.2\% of census targets for both CBGs. The residuals do not display any clear spatial pattern, suggesting that the allocation does not introduce systematic spatial bias that could affect downstream hazard or impact analyses.

\subsubsection{Demographic Marginal Distribution Alignment} 

\begin{figure}[htbp]
    \centering
    \includegraphics[width=0.95\textwidth]{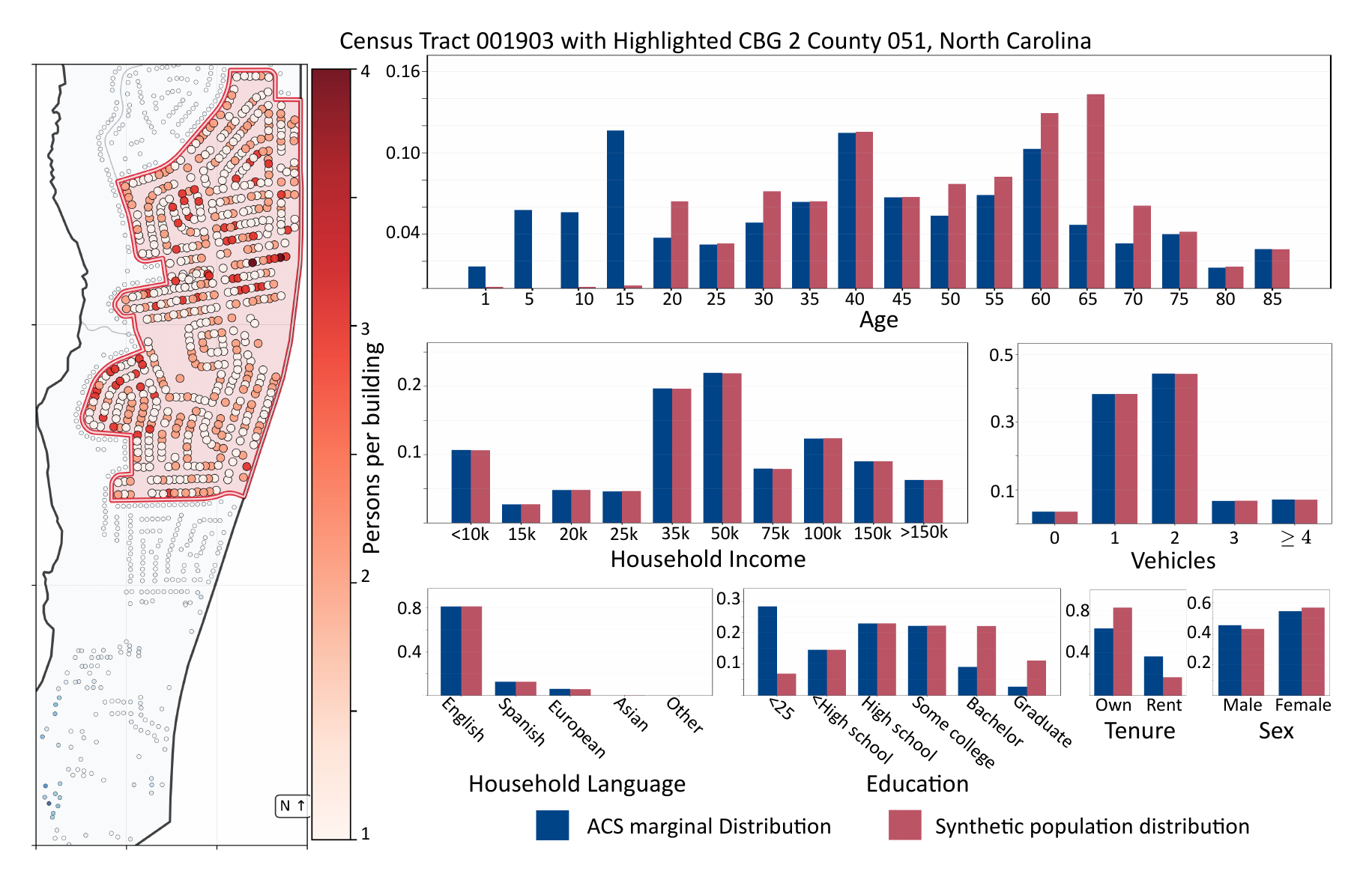}
    \caption{Comparison of marginal distributions between synthetic allocation and ACS benchmark for CBG 370510019032.}
    \label{fig:marginal_distributions}
\end{figure}

A key feature of a high-quality joint housing-household inventory is that its demographic distributions realistically resemble those of the real population, while remaining fully synthetic. To illustrate this alignment, we plot the marginal distributions of household- and person-level attributes for CBG 370510019032 in Figure~\ref{fig:marginal_distributions}, which compares the synthetic population with demographic statistics published by the ACS. This CBG includes approximately 478 households and represents a typical suburban neighborhood with a mix of housing types. We compare the demographics of the allocated synthetic population with ground-truth marginal distributions from the ACS 5-year estimates.

We use ACS 5-year rather than 1-year estimates because CBG-level demographic data are only available through 5-year aggregations; 1-year estimates are not published for geographies with fewer than 65,000 residents \citep{census_education_factsheet_2021}. While 5-year data reflect averaged rather than single-year conditions, they provide the most detailed and reliable demographic information at this spatial granularity. Because the synthetic population reports continuous values for some attributes (e.g., household income), we discretize these into standard census brackets and aggregate detailed educational categories into the broader ACS groupings to enable direct comparison.

At the household level, the synthetic and ACS distributions show strong agreement. The household income distribution (HINCP) captures both the primary peak in the \$35,000-\$50,000 range and a secondary peak in the \$75,000-\$100,000 range. Household language (HHL) closely matches census statistics, with about 80\% English-only households and the remaining 20\% speaking Spanish or other languages. Vehicle availability (VEH) also aligns well with ACS data, correctly reproducing the dominant two-vehicle households (about 45\%), as well as car-free households (3\%) and those with three or more vehicles (15\%). Housing tenure (TEN) shows somewhat more variation, but still achieves reasonable alignment with observed homeownership rates (RMSE = 0.19).

Person-level results further support the framework's validity. Educational attainment (SCHL) shows larger but still acceptable differences (RMSE = 0.109) under the ACS age-25+ benchmark, reproducing mid-level education well while slightly underestimating individuals with less than a high school education and overestimating those with advanced degrees. This is consistent with the conservative preprocessing rule in Section~\ref{sec:synthetic_household_data_generation_model}: individuals younger than 25 remain in the joint model, but their \texttt{SCHL} values are carried as a separate categorical state rather than forced into adult attainment categories. The sex distribution (SEX) is reproduced with high accuracy (RMSE = 0.024). The allocation still under-generates individuals younger than 20 and overestimates the elderly population, suggesting that age structure remains harder to match than sex or adult educational attainment under the current variable set. Future work could reduce this residual bias by incorporating richer youth-specific covariates or auxiliary data sources describing school progression more directly.

\subsubsection{Regional Joint Inventory Performance Assessment}

The overall performance of the joint housing-household inventory is assessed and illustrated in Table~\ref{tab:cbg_table} using a set of randomly selected CBGs. ``Housing Units'' denotes the number of housing units in the building inventory, and ``Household'' records the number of households assigned to housing units. 

For each building, household allocations to housing units are required to satisfy capacity and demographic constraints (e.g., limits on the number of older adults). Let building index $b$ and constraint $c$ define a non-negative exceedance $e_{b,c}$, which is zero when the constraint is satisfied. We also report the number of buildings that violate capacity constraints (``Flagged bldgs''). In addition, we report the proportion of buildings that receive household assignments satisfying the building capacity constraints (``Feasible bldg'').

For a household or person attribute $A$ with categories $k=1..K$, let $p^{gt}_{A,k}$ and $p^{alloc}_{A,k}$ denote the ground-truth and allocated proportions within a CBG, respectively. We compute
 \begin{equation}
\mathrm{RMSE}(A)=\sqrt{\tfrac{1}{K}\sum_k (p^{alloc}_{A,k}-p^{gt}_{A,k})^2}
\end{equation}
and then average RMSE values across household attributes (tenure, vehicles owned, Household income, and language) to obtain ``RMSE (Household)'', and across person attributes (age, sex, school attainment) to obtain ``RMSE (Person)''.

\begin{table}[t]
    \centering
    \small
    \caption{Joint housing-household inventory performance at randomly selected CBGs}
    \resizebox{\columnwidth}{!}{%
    \begin{tabular}{lccccccc}
    \toprule
    CBG (FIPS)                                                      & \multicolumn{1}{c}{\begin{tabular}[c]{@{}c@{}}Housing\\ Units\end{tabular}} & \multicolumn{1}{c}{Household} & \multicolumn{1}{c}{\begin{tabular}[c]{@{}c@{}}Feasible \\ bldgs(\%)\end{tabular}} & \multicolumn{1}{c}{\begin{tabular}[c]{@{}c@{}}Flagged \\ bldgs\end{tabular}} & \multicolumn{1}{c}{\begin{tabular}[c]{@{}c@{}}RMSE \\ (HH)\end{tabular}} & \multicolumn{1}{c}{\begin{tabular}[c]{@{}c@{}}RMSE \\ (Person)\end{tabular}} & \multicolumn{1}{c}{\begin{tabular}[c]{@{}c@{}}Median \\ compatibility\end{tabular}} \\ 
    \midrule
    371010402031         & 2393          & 2391     & 0.9992        & 2             & 0.1407    & 0.1012        & 0.6868       \\
    371950015002         & 894           & 890      & 0.9989        & 1             & 0.0972    & 0.0858        & 0.7192       \\
    $\cdots$             & $\cdots$      & $\cdots$ & $\cdots$      & $\cdots$      & $\cdots$  & $\cdots$      & $\cdots$     \\
    370179502001         & 570           & 566      & 0.9982        & 1             & 0.1328    & 0.1025        & 0.6774       \\
    370179504004         & 449           & 449      & 0.9978        & 1             & 0.0695    & 0.0682        & 0.7582       \\
    370510002003         & 238           & 238      & 0.9958        & 1             & 0.0571    & 0.0942        & 0.7571       \\ \midrule
    \multicolumn{1}{c}{\begin{tabular}[c]{@{}c@{}}All  \\ (mean $\pm$  sd)\end{tabular}} & \multicolumn{1}{c}{\begin{tabular}[c]{@{}c@{}}576.7 \\ $\pm$  332.7\end{tabular}} & \multicolumn{1}{c}{\begin{tabular}[c]{@{}c@{}}549.0 \\ $\pm$  315.9\end{tabular}} & \multicolumn{1}{c}{\begin{tabular}[c]{@{}c@{}}0.8825 \\ $\pm$  0.1374\end{tabular}} & \multicolumn{1}{c}{\begin{tabular}[c]{@{}c@{}}64.5 \\ $\pm$  81.4\end{tabular}}   & \multicolumn{1}{c}{\begin{tabular}[c]{@{}c@{}}0.1313 \\ $\pm$  0.0834\end{tabular}} & \multicolumn{1}{c}{\begin{tabular}[c]{@{}c@{}}0.0798 \\ $\pm$  0.0421\end{tabular}} & \multicolumn{1}{c}{\begin{tabular}[c]{@{}c@{}}0.7664 \\ $\pm$  0.0729\end{tabular}} \\
    \bottomrule
    \end{tabular}
    }
    \label{tab:cbg_table}
\end{table}

The contrastive learning model assigns a compatibility score to each candidate housing-household unit pair. For realized assignments, we summarize the median compatibility score between housing units and assigned households at the CBG level. 

The results show that our framework performs well across the region, producing realistic and reliable housing-household assignments. On average, 88.3\% of households are placed in housing units within residential buildings that fully meet all constraints, such as capacity limits and age-related requirements. Only a small number of buildings (about 65) show any violations, indicating that the allocation process effectively adheres to these rules. The produced joint housing-household inventory also closely matches real-world demographic patterns. The overall errors in household and person characteristics are low, meaning the assigned populations closely reflect census data at the neighborhood level. In addition, the high median compatibility score (0.77) shows that households are generally matched to suitable housing units, as identified by the model. This strong performance holds across a wide range of neighborhoods, from dense urban areas to smaller communities, demonstrating that the approach is adaptable.

\begin{figure}[htbp]
    \centering
    \includegraphics[width=0.95\textwidth]{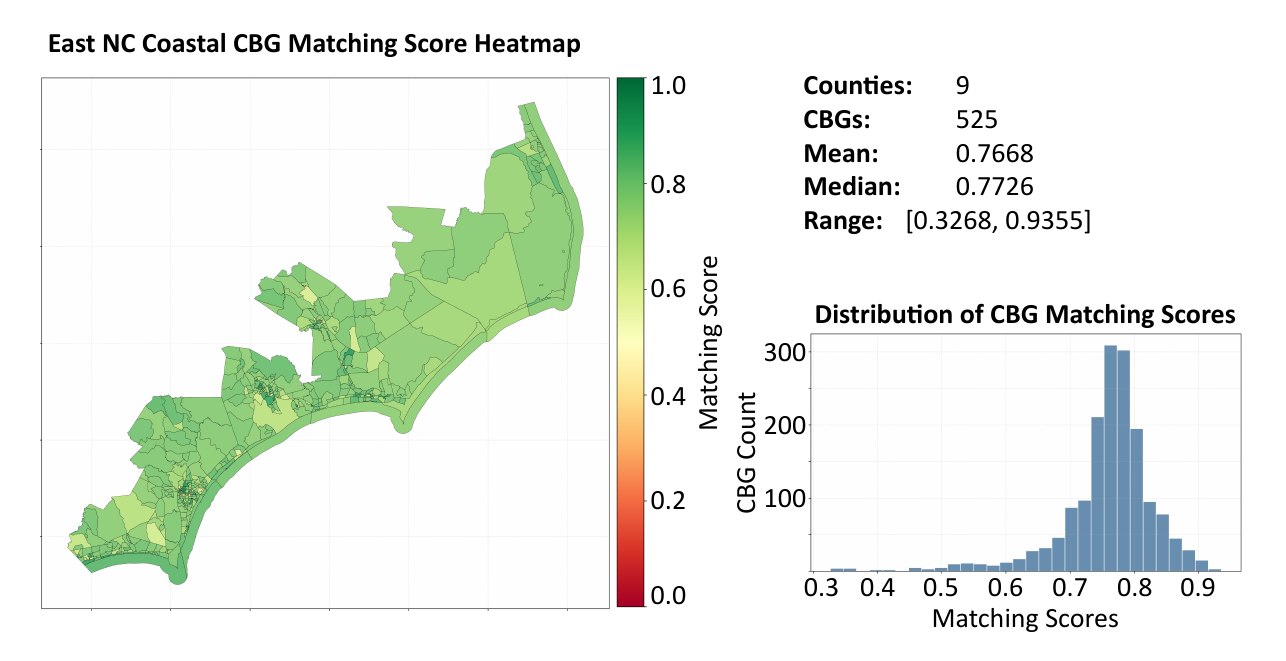}
    \caption{Distribution of housing-household compatibility scores in the final joint inventory.}
    \label{fig:regional_matching_quality}
\end{figure}

Finally, we construct a regional joint housing-household inventory for all counties along the North Carolina coast. The results demonstrate that the proposed method performs consistently across diverse geographic contexts, including urban, suburban, and rural coastal communities. Figure~\ref{fig:regional_matching_quality} presents the mean housing-household matching scores for the nine coastal counties. All counties achieve mean scores above 0.75, indicating that the deep contrastive learning model reliably identifies high-quality housing-household matches. This regional evaluation confirms that the framework maintains strong allocation performance across heterogeneous coastal settings. The observed spatial consistency suggests that the approach scales effectively from densely populated urban counties, such as New Hanover County with 118 CBGs, to sparsely populated rural counties, such as Hyde County with only 4 CBGs. Maintaining comparable matching scores across these settings highlights the robustness of the framework to differences in sample size and geographic diversity.

\section{Discussion}

The joint inventory enables analyses that neither structure-only nor population-only datasets can support, because each building carries a linked household with known demographic attributes. In disaster risk assessment, loss estimation models can produce building-specific projections of displacement duration, shelter demand, and recovery cost that reflect resident income and composition, rather than treating all occupants of a given building type identically. Emergency managers can overlay the inventory on evacuation networks to identify where mobility-limited or linguistically isolated households concentrate relative to building types, enabling resource pre-positioning at a resolution that is finer than a tract. Coupling household income and tenure with structural characteristics (building age, number of units, property value) further supports parcel-level identification of cost-burdened or overcrowded households, informing zoning and affordability interventions that would otherwise rely on aggregate neighborhood statistics. Energy agencies can direct weatherization and utility-assistance programs toward buildings where low-income households occupy energy-inefficient structures, improving the targeting precision of retrofit subsidies. In public health, linking household age structure and housing conditions to specific buildings supports spatially targeted interventions for heat-related illness, lead exposure, and indoor air quality, where building envelope and occupant profile jointly determine risk. Across these domains, the inventory converts analyses previously feasible only at aggregate geographic scales into building-level assessments.

Although the empirical validation here focuses on nine coastal counties in North Carolina, the framework is supported by three functions that exist in many national statistical systems: a household microdata source with linked demographic and dwelling attributes, a building or address register with residential-unit characteristics, and small-area control totals for allocation. In the United States, ACS PUMS, the NSI, and ACS small-area tabulations can be used. Comparable inputs can be found elsewhere. However, access conditions vary. England and Wales provide Census 2021 microdata and geographies through the ONS \citep{ons_microdata_2024, ons_census_geographies_2022}; Statistics Canada offers housing-related public-use microdata and dissemination-area tabulations \citep{ statcan_chs_pumf_2021, statcan_census_dictionary_2021}; and the Netherlands combines CBS microdata and open neighbourhood statistics with the BAG address-and-building register \citep{cbs_microdata_2025, cbs_open_data_2025, cbs_neighbourhood_data_2025, kadaster_bag_linked_data_2025}. The microdata component is not always open access and may be available only through safeguarded research environments. Therefore, the ACS-to-NSI implementation presented here should be understood as one U.S. demonstration of a broader workflow rather than a country-specific design.

This paper directly delivers a usable joint inventory for coastal Eastern North Carolina. Beyond this case study, the same three-module workflow can be reapplied elsewhere. Researchers fit the synthetic population module to the local household distribution, re-estimate the compatibility model from locally observed co-occurrence patterns when available, and run the optimization with the local building inventory, control totals, and housing definitions. In data-sparse settings, compatibility features and allocation constraints must be simplified to reflect the available information. The overall architecture remains the same across study areas, while the fitted models, variable mappings, and allocation constraints are region-specific. The three modules can also be further improved with state-of-the-art models and proprietary datasets. 


\section{Limitations and Future Work}

The allocation model assigns each household to exactly one housing unit, and we do not capture the case where a single household owns multiple other properties. The ACS defines a household as the persons who occupy a housing unit as their usual place of residence \citep{acs_subject_definitions_2022}, and each individual is recorded at exactly one address regardless of how many properties they may own \citep{hvs_definitions_2020}. Additional properties enter the data independently: if occupied by tenants, they appear as separate renter-occupied units with their own household; if unoccupied, they are classified as vacant under one of several ACS vacancy categories (e.g., seasonal, for rent, or other vacant). The NSI follows the same residence-based logic, distributing census-block population totals across structures in proportion to housing units rather than ownership \citep{nsi_tech_2022}. Because both inputs record people at their place of residence rather than at properties they own, the one-to-one household-unit mapping in our model mirrors the data-generating process. Multi-property ownership manifests in the data either as separate renter-occupied units (each with its own tenant household) or as vacant units regulated by the vacancy-rate constraints in our optimization formulation.

The framework depends on the quality and completeness of its input data sources. Where the NSI omits residential structures or misestimates capacity, the allocation may assign more residents than the physical stock can accommodate. Demographic alignment is further constrained by the use of ACS 5-year estimates, which represent period-averaged conditions and may obscure short-term dynamics such as seasonal migration or rapid neighborhood turnover. As with other deep generative approaches, the synthetic population module may occasionally produce rare or atypical household compositions. Thus, future work could add a gating mechanism to filter implausible cases before allocation. The contrastive learning model achieves high held-out accuracy, but its limitations are inherited from the underlying ACS sample. As discussed in \citet{qian2025dcl}, the public microdata represent only a limited sample of the full housing and household population, leaving geographic coverage gaps. Furthermore, because the data contain observed co-occurrences but no explicit negatives or a fully labeled test set, the DCL framework relies on pseudo-labels, cluster-augmented pairs, and synthetic ground truth, which mitigate but cannot eliminate residual false negatives and noise.

A related constraint is the use of property value as the sole monetary housing attribute. Section~3.2.1 shows that property value and gross rent remain strongly rank-correlated after conditioning on the structural attributes used in matching, supporting its use as an ordinal proxy. However, the model cannot represent rental-market-specific variation where rent and asset value diverge more sharply.

In addition, direct validation against observed household-to-address records is outside the scope of the present study because ACS public-use files intentionally remove identifying information. Census responses that could identify a specific individual or address cannot be publicly released \citep{title13_confidentiality_2026}. We therefore restrict the current framework to public-use microdata, the NSI, and ACS aggregate tables. Future work could extend evaluation through secure and private data partnerships that provide privacy-protected administrative records or field-survey linkages.

Furthermore, the sequential, modular architecture provides no feedback mechanism from the MILP allocation stage back to the diffusion-based generative model when demographic edge cases are underrepresented in the candidate pool. Coupling the two through gradient-based feedback would require a differentiable path from the integer program back through the discrete sampling process, which remains an open problem in combinatorial optimization \citep{bengio2021machine}. The oversampling strategy described in Section~\ref{sec:oversampling_rationale} mitigates this in practice but does not eliminate it. An end-to-end differentiable architecture would, in principle, allow such feedback, but it requires a building-level household-to-unit training dataset, which is precisely the motivation and the product of this research. Every existing method in the population synthesis literature, from iterative proportional fitting \citep{ye2024enhancing} to Bayesian networks and deep generative models \citep{chapuis2022generation}, follows the same modular pattern, so the sequential design is consistent with the current state of the field.

Finally, the oversampled population candidate pool preserves feasibility under demographic and capacity constraints but enlarges the MILP and increases solve time, especially in dense block groups. The stage limits reported in Section~4 are therefore upper bounds per block group, not for a full tract or the entire study region. Across all runs, the median solve time was approximately 1.2 hours, and most CBGs finished in 0.9-2.0 hours on a workstation equipped with an AMD Ryzen Threadripper PRO 7975WX CPU and 250~GiB RAM. Larger instances, such as block groups with more than 1,700 housing units, are more demanding and thus were moved to TACC's Stampede3 \texttt{nvdimm} queue (Intel Xeon Platinum 8380 CPUs, 4~TB NVDIMM memory). The expanded memory allows these MILP instances to remain resident, although it does not provide the same CPU-side advantage as the local workstation. On lower-memory machines, these instances would run longer or fail under the present implementation. Further improvements can apply tighter candidate-screening rules, decomposition strategies, or specialized optimization algorithms to reduce runtime without sacrificing demographic fidelity or compatibility quality.

\section{Conclusion}

This paper presents a framework (Figure 1) for constructing a joint housing-household inventory that explicitly links individuals within households to compatible housing units while preserving realistic population density and demographic distributions. By integrating high-resolution building data from the NSI with a synthetic population generated from ACS PUMS, and by leveraging a deep contrastive learning model to infer housing-household compatibility, the proposed approach bridges a critical knowledge and data gap between structure-focused and population-oriented datasets used in existing disaster impact assessment and resilience analysis.

The framework's effectiveness is supported by strong performance across its three core components. The synthetic population module (Figure 2) generates households that resemble census microdata, producing realistic synthetic individuals and households that are statistically indistinguishable from ACS PUMS (Column Shape = 0.9985, Column Pair Trend = 0.9946, C2ST = 0.9806) while preserving complex intra-household relationships (Table 2). The deep contrastive learning model (Figure 3) effectively captures compatibility between household characteristics and housing attributes, achieving high predictive accuracy (92.7\%) and a Matthews correlation coefficient (MCC) of 0.854 on held-out validation data (Table 3). Finally, the hierarchical allocation procedure (Sec. 3.3) balances feasibility and demographic realism, assigning 98.8\% of households to buildings that satisfy capacity and demographic constraints while maintaining close alignment with census-based population distributions at the census block group (CBG) level.

The joint inventory (Figure~4) is evaluated across multiple spatial scales. At the building level, spatial analyses show that the synthetic population preserves observed population density patterns without introducing systematic spatial bias (Figure 5). At the neighborhood level, the joint housing-household inventory closely reproduces census-based marginal distributions of both household- and person-level attributes while maintaining high compliance with building-level constraints (Figure 6). At the regional scale, consistent housing-household compatibility quality across all coastal counties of North Carolina (Table 4, Figure 7), spanning urban, suburban, and rural contexts, demonstrates the robustness and scalability of the framework.


\section*{Acknowledgments}
The authors would like to acknowledge funding support from the National Science Foundation \#2443784 and \#2209190. Any opinions, conclusions, and recommendations expressed in this research are those of the authors and do not necessarily reflect the views of the funding agencies. The authors would also like to thank the editor and the anonymous reviewers for their constructive comments and valuable insights to improve the quality of the article. 

\bibliography{reference}
\bibliographystyle{citation_style_bcu}


\end{document}